\documentclass{aa}  

\usepackage{graphicx}
\usepackage{txfonts}
\usepackage{graphicx}	
\usepackage{amsmath}	
\usepackage{xcolor, soul}

\newcommand{\angstrom}{\mbox{\normalfont\AA $~$}}

\begin{document}

	\title{Eddington Ratios of Dust-obscured Quasars at $z \lesssim 1$:\\
			Evidence Supporting Dust-obscured Quasars as Young Quasars}
	\titlerunning{Eddington Ratios of Dust-obscured Quasars at $z \lesssim 1$}
	
	\author{
		Dohyeong Kim\inst{1}, 
		Yongjung Kim\inst{2}, 
		Myungshin Im\inst{3},
		Eilat Glikman\inst{4},
		Minjin Kim\inst{5},
		Tanya Urrutia\inst{6},
		and Gu Lim\inst{1}
	}
	\institute{
		Department of Earth Sciences, Pusan National University, Busan 46241, Republic of Korea \and
		Korea Astronomy and Space Science Institute, Daejeon 34055, Republic of Korea \and
		SNU Astronomy Research Center (SNU ARC), Astronomy Program, Dept. of Physics \& Astronomy, Seoul National University, Seoul 08826, Republic of Korea \and
		Department of Physics, Middlebury College, Middlebury, VT 05753, USA \and
		Department of Astronomy and Atmospheric Sciences, College of Natural Sciences, Kyungpook National University, Daegu 41566, Republic of Korea \and
		Leibniz-Institut für Astrophysik, Potsdam (AIP), An der Sternwarte 16, 14482 Potsdam, Germany \\
		\email{dh.dr2kim@gmail.com; yjkim.ast@gmail.com}}
	\authorrunning{Kim et al. }
	\date{--; --}
	
	
	\abstract
	{Dust-obscured quasars have been suspected as the intermediate stage galaxies between
	merger-driven star-forming galaxies and unobscured quasars.
	This merger-driven galaxy evolution scenario suggests that 
	dust-obscured quasars exhibit higher Eddington ratios ($\lambda_{\rm Edd}$) than those of unobscured quasars.
	However, their high dust obscuration poses challenges to accurately measuring their $\lambda_{\rm Edd}$
	using commonly employed bolometric luminosity ($L_{\rm bol}$) and black hole (BH) mass ($M_{\rm BH}$) estimators
	based on the ultraviolet (UV) or optical luminosity. 
	Recently, \cite{kim23} established new estimators for $L_{\rm bol}$ and $M_{\rm BH}$ based on mid-infrared (MIR) continuum luminosity ($L_{\rm MIR}$), 
	which are less affected by dust obscuration.
	These estimators enable the study of a large number of dust-obscured quasars across a wide redshift range.
	In this study, we measure the $\lambda_{\rm Edd}$ values of 30 dust-obscured quasars at $z \lesssim 1$,
	the largest sample size to date, using the $L_{\rm MIR}$-based $L_{\rm bol}$ and $M_{\rm BH}$ estimators.
	Our findings reveal that dust-obscured quasars exhibit significantly higher $\lambda_{\rm Edd}$ values compared to unobscured quasars.
	Moreover, we confirm that the enhanced $\lambda_{\rm Edd}$ values of dust-obscured quasars maintain consistency across the redshift span of 0 to 1.
	Our results strongly support the picture that dust-obscured quasars are 
	in the earlier stage than unobscured quasars in the merger-driven galaxy evolutionary track.
	}
	
	\keywords{Galaxies: active -- (Galaxies:) quasars: general -- (Galaxies:) quasars: supermassive black holes -- 
		(Galaxies:) quasars: emission lines -- Galaxy: evolution -- Infrared: galaxies}
	
	\maketitle
	
	%
	
	\section{Introduction}\label{sec:intro}
	Quasars are one of the most energetic and luminous objects in the universe,
	and they emit the enormous energy in all wavelengths from gamma-ray to radio.
	The energy of quasars is powered by the accretion of surrounded materials onto supermassive black holes (SMBHs).
	These SMBHs are found at the centers of massive spheroidal galaxies,
	and the BH masses have correlations with several properties of host galaxies
	(e.g., \citealt{ferrarese00,gebhardt00}).
	
	To date, almost million quasars have been found based on X-ray, UV, optical, and radio surveys
	\citep{grazian00,becker01,anderson03,croom04,risaliti05,schneider05,veron-cetty06,young09,paris14,kim15c,kim19,kim20b,lyke20,shin20,kim22b,shin22}.
	However, some previous studies (e.g., \citealt{comastri01,tozzi06,polletta08}) reported that 
	soft X-ray, UV, and optical quasar surveys may overlook a substantial number of quasars,
	which have red colors caused by the dust extinction from intervening dust and gas
	in their host galaxies \citep{webster95,cutri02} or our galaxy \citep{im07,lee08}.
	In this paper, we refer to the quasars with the dust extinction from their host galaxies as dust-obscured quasars.
	Some previous studies (e.g., \citealt{polletta08}) expected that
	these dust-obscured quasars could make up $\sim$50\,\% of the whole quasar population,
	and even the optical quasar surveys include a considerable number of dust-obscured quasars (e.g., $\sim$15\,\%; \citealt{kim23}).
	
	Dust-obscured quasars have been found through various methods,
	but the use of their red colors is a notably useful technique.
	Especially, the colors in near-infrared (NIR; e.g., $J-K>2$\,mag; \citealt{cutri01,cutri02})
	and optical through NIR (e.g., $R-K > 5$\,mag and $J-K>1.3$\,mag; \citealt{glikman07,urrutia09}) have been widely used. 
	The quasars found by these red colors are called red quasars,
	and using these red colors proves to be an effective way to find dust-obscured quasars
	since the red colors of most red quasars originate from the dust extinction \citep{kim18a}.
	However, \cite{kim18b} claimed that using only NIR color is not effective in selecting dust-obscured quasars.
	
	\begin{figure*}
	\centering
	\includegraphics[width=\textwidth]{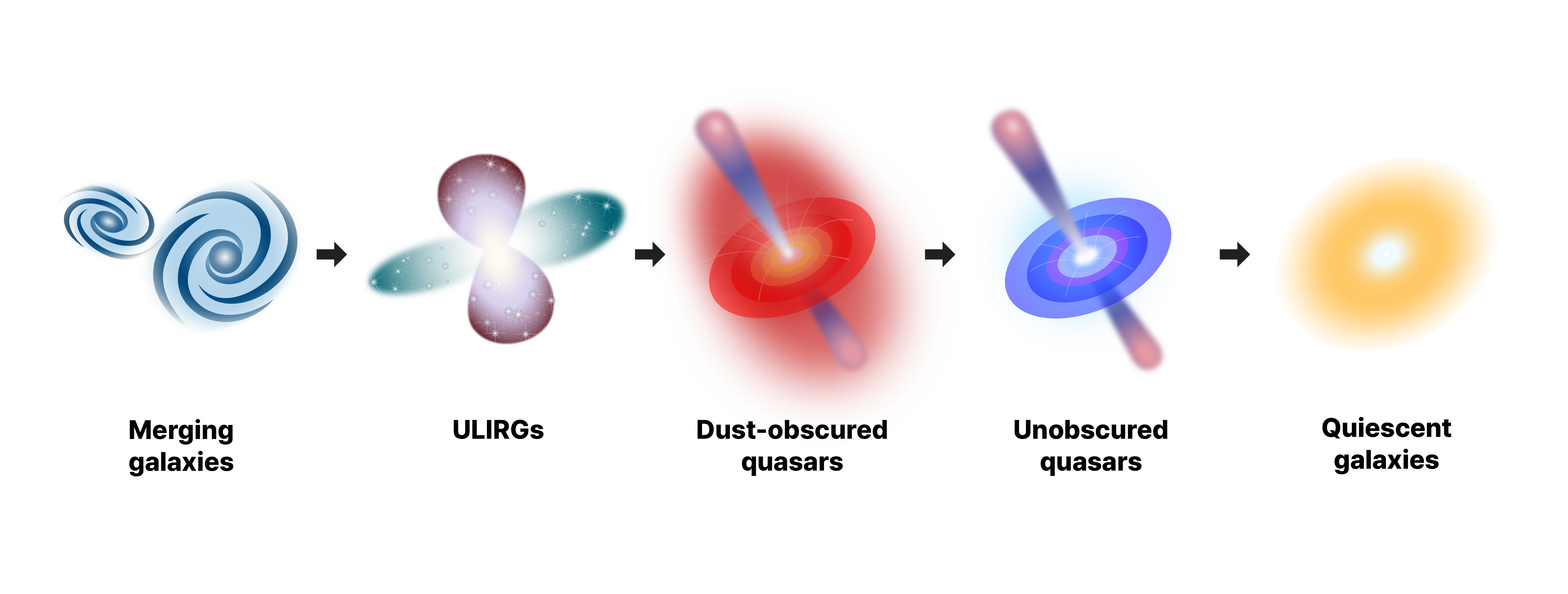}\\
	\caption{
		Schematic outline of merger-driven galaxy evolution scenario.
		\label{fig:Gal_evol}}
	\end{figure*}	
	
	Although our knowledge about dust-obscured quasars is still incomplete,
	the emergence of dust-obscured quasars is commonly associated with
	a merger-driven galaxy evolution scenario, extensively explored in previous simulation studies \citep{menci04,hopkins05,hopkins06,hopkins08}.
	In this merger-driven galaxy evolution scenario, 
	major mergers between gas-rich galaxies can trigger both star-formation and BH activities,
	and these phase galaxies are often seen as ultra-luminous infrared galaxies (ULIRGs; \citealt{sanders88,sanders96}).
	After this phase, these galaxies can have merging features and star-formation activities.	
	Furthermore, since gas flows onto the SMBHs are related to
	galaxy-scale disturbances caused by galaxy merger-driven torques \citep{dimatteo05,hopkins08,alexander12},
	it is expected that their BH activities increase dramatically,
	resulting in the emitting enormous amounts of energy with a high $\lambda_{\rm Edd}$.
	However, their BH activities are still obscured owing to the remaining dust and gas in their host galaxies.
	Finally, these galaxies evolve to normal unobscured quasars
	after sweeping away the remaining dust and gas by quasar-driven winds.
	Figure \ref{fig:Gal_evol} shows a schematic outline of this merger-driven galaxy evolution scenario.
	
	In this scenario, it is believed that dust-obscured quasars can occur
	as the intermediate stage galaxies between the merger-driven star-forming galaxies and unobscured quasars.
	Several theoretical studies (e.g., \citealt{hopkins08}) have suggested that 
	dust-obscured quasars exhibit high BH accretion rates, which appear to be prevalent.
	This scenario has been supported by several pieces of observational evidence.
	For example, dust-obscured quasars have 
	(i) high $\lambda_{\rm Edd}$ values \citep{urrutia12,kim15a,kim18b,kim24},
	(ii) high fractions of merging features in their host galaxies \citep{urrutia08,glikman15},
	(iii) dusty red colors \citep{kim18a},
	(iv) merging SMBH systems \citep{kim20},
	and (v) enhanced star-formation activities \citep{georgakakis09}.
	
	Note, however, that even in the merger-driven galaxy evolution scenario,
	the properties of dust-obscured quasars remain controversial.
	Recently, \cite{lansbury20} showed that the dust-obscured quasar phase lasts a few $10^5$\,years,
	which is significantly shorter than quasar lifetime of $\sim$$10^7$\,years (e.g., \citealt{hopkins06,khrykin21}).
	Considering this significantly short dusty phase, some unobscured quasars might have SMBHs as active as those in dust-obscured quasars.
	However, these lifetimes are sensitive to several properties (e.g., $L_{\rm bol}$ and stellar mass; \citealt{hopkins06}),
	and hence more extensive observational studies on the quasar lifetimes are necessary to obtain a conclusive understanding
	of the BH activities of dust-obscured quasars.
	Moreover, some recent studies (e.g., \citealt{jun21,glikman24}) showed that outflows slowly sweeping away the dust and gas in host galaxies
	could result in luminous dust-obscured quasars with enhanced BH activities.
	
	Alternative explanation for dust-obscured quasars is that they are obscured by a dust torus.
	The unification model suggests that the differences in emission lines between type 1 and 2 active galactic nuclei
	are due to the viewing angle \citep{antonucci93,urry95}.
	Some previous studies (e.g., \citealt{wilkes02,rose13,ananna22a,ananna22b}) proposed that the red colors can be
	a result of moderate viewing angle in the unification model,
	where dust torus blocks the photon from accretion disk and broad line region (BLR),
	instead of the dust and gas in their host galaxies.
	In this scenario, dust-obscured quasars appear solely due to the viewing angle,
	so their BH accretion rates should not differ from those of unobscured quasars.
	Even when considering the dust torus evolution with BH accretion rates \citep{ananna22a,ananna22b},
	a significant number of dust-obscured quasars are expected to exhibit lower BH accretion rates than unobscured quasars.
	
	Another possible explanation for dust-obscured quasars is that strong outflows play a crucial role 
	in transporting the obscuring medium \citep{calistro21}. 
	Several observational studies showed that dust-obscured quasars exhibit strong outflows (e.g., \citealt{lansbury20,stacey22,glikman24}),
	which support this explanation. 
	Note, however, that strong outflows are also expected to occur in the intermediate stage galaxies of the merger-driven galaxy evolution scenario,
	indicating that it cannot be considered a completely separate explanation from the merger-driven galaxy evolution scenario.
	
	Furthermore, some previous studies (e.g., \citealt{puchnarewicz98,whiting01,rose14})
	suggested that red quasars have intrinsic red colors, instead of the dust extinction.
	Additionally, \cite{georgantopoulos23} demonstrated that red quasars reside in the galaxies
	located between the young and old galaxy populations,
	in contrast to expectations from the merger-driven galaxy evolution scenario.
	
	
	One of the ways to confirm if dust-obscured quasars are the intermediate population
	under the merger-driven galaxy evolution scenario
	is to examine whether dust-obscured quasars have high $\lambda_{\rm Edd}$ values.
	However, the $\lambda_{\rm Edd}$ values of dust-obscured quasars remain a topic of debate due to that
	most of bolometric luminosity and BH mass estimators are established based on UV or optical luminosity
	\citep{kaspi00,vestergaard02,mclure04,greene05},
	where the UV- and optical-based estimators are easily affected by the dust extinction.
	For example, if a dust-obscured quasar is obscured by a color excess of $E(B-V)=2$\,mag,
	its fluxes at 1450\,\angstrom and 5100\,\angstrom are suppressed by
	factors of $4.3 \times 10^{6}$ and 500, respectively,
	which is obtained from a reddening law \citep{fitzpatrick99} with $R_{V}=3.1$ \citep{weingartner01}.
	Furthermore, to determine the $E(B-V)$ values for dust-obscured quasars is also uncertain;
	the differences in $E(B-V)$ estimates can vary as much as 1--2\,mag, 
	depending on the method used \citep{glikman07,urrutia12,kim18b}.
	If a dust-obscured quasar with a color excess of $E(B-V)=1$\,mag is measured as having $E(B-V)=2$\,mag, 
	then its fluxes at 1450\,\angstrom and 5100\,\angstrom
	are overestimated by $\sim$2000 and $\sim$20 times, respectively, after correcting the dust extinction.
	Recently, \cite{calistro21} measured the $\lambda_{\rm Edd}$ values of $\sim$300 dust-obscured quasars using UV and optical luminosities, 
	and found no significant difference from those of unobscured quasars.
	However, \cite{calistro21} used UV and optical luminosities to measure the BH properties,
	which are sensitive to accuracy in $E(B-V)$ measurements.
	
	To overcome the limitations of the UV- and optical-based estimators,
	several infrared (IR)-based estimators have been established.
	If a dust-obscured quasar with $E(B-V)=1$\,mag is incorrectly measured as $E(B-V)=2$\,mag,
	then the dust extinction corrected fluxes of P$\alpha$ (1.87\,$\mu$m) and Br$\alpha$ (4.05\,$\mu$m) lines
	would be overestimated by only a factor of  1.48 and 1.16, respectively.
	For this advantage, Paschen and Brackett hydrogen line based estimators were derived (e.g., \citealt{kim10,kim15b,kim22}).
	However, despite the advantage, these IR-based estimators need IR spectroscopic observation,
	and applying them to a large number of dust-obscured quasars has been a tough task due to the difficulty in obtaining IR spectroscopic data.
	Consequently, while several previous studies \citep{kim15a,kim18a,kim18b} showed that
	the $\lambda_{\rm Edd}$ values of dust-obscured quasars are higher than those of unobscured quasars using these IR-based estimators,
	they used the limited sample size ($\lesssim$15),
	yielding the results are (i) not statistically robust and (ii) not confirmed for a wide redshift range.
	
	Recently, \cite{kim23} established bolometric luminosity and BH mass estimators based on $L_{\rm MIR}$
	using 129 unobscured ($E(B-V) < 0.1$) Sloan Digital Sky Survey (SDSS; \citealt{york00}) quasars at $z \lesssim 0.5$.
	The $L_{\rm MIR}$ values can be derived via spectral energy distribution (SED) fitting
	using various photometric data, such as
	SDSS (\citealt{york00}), Two Micron All-Sky Survey (2MASS; \citealt{skrutskie06}),
	and Wide-field Infrared Survey Explorer (WISE; \citealt{wright10}).
	The bolometric luminosity can be measured from the $L_{\rm MIR}$ with an rms scatter of $\sim$0.1\,dex,
	while the BH mass can be derived with an rms scatter of $\lesssim$0.2\,dex
	using the $L_{\rm MIR}$ with the FWHM of Balmer lines.
	Since the measurement of the FWHM of Balmer lines is not significantly affected by the dust extinction,
	the $L_{\rm MIR}$-based estimators allow for the determination of $L_{\rm bol}$	and $M_{\rm BH}$
	without the need for IR spectroscopic data and relatively unaffected by the dust extinction.
	Hence, these $L_{\rm MIR}$-based estimators can be applied for a large number of dust-obscured quasars,
	and they are expected to yield the statistically robust result of the $\lambda_{\rm Edd}$ values of dust-obscured quasars.
	
	In this paper, we measure the $\lambda_{\rm Edd}$ values of 30 dust-obscured quasars at $z \lesssim 1$
	using the $L_{\rm MIR}$-based estimators \citep{kim23},
	and compare them to unobscured quasars selected from the SDSS quasars.
	Throughout this work, we use a standard $\Lambda$CDM model of
	$H_{0}=70\,{\rm km\,s^{-1}}$\,Mpc$^{-1}$, $\Omega_{m}=0.3$, and $\Omega_{\Lambda}=0.7$.
	This model has been supported by observational studies 
	in the past decades (e.g. \citealt{im97,planck16}).
	
	\section{Sample}\label{sec:sample}
	
	Our sample is drawn from 120 dust-obscured quasars listed in \cite{glikman12},
	and these dust-obscured quasars were selected via the following procedures.
	First, they selected dust-obscured quasar candidates using
	(i) detections in NIR (2MASS Point Source Catalog [PSC]; \citealt{cutri03}) 
	and radio (Faint Images of the Radio Sky at Twenty-centimeters [FIRST]; \citealt{becker95});
	and (ii) red colors in NIR ($J-K>1.7$\,mag) and optical through NIR ($R-K > 4$\,mag).
	Through this process, a total of 395 dust-obscured quasar candidates were selected.
	Second, \cite{glikman12} obtained the optical and/or NIR spectra of $\sim$300 dust-obscured quasar candidates
	for (i) object identification; (ii) redshift determination; and (iii) dust reddening determination.
	Finally, of the $\sim$300 candidates, they found 120 dust-obscured quasars having $E(B-V) > 0.1$,
	and provide their optical and/or NIR spectra.
	
	Among the 120 dust-obscured quasars, 
	we select 30 dust-obscured quasars at $z \lesssim 1$ by following conventions.
	First, broad component of H$\beta$ or H$\alpha$ line is 
	broader than 2000\,$\rm km~s^{-1}$ (see Section \ref{sec:Line_fit} for more details).
	Second, $E(B-V)$ measured from spectral energy distribution (SED) fitting is greater than 0.1,
	and the details of the SED fitting procedure are described in Section \ref{sec:SED_fit}.
	Third, quasar fractions at 3.4\,$\mu$m and 4.6\,$\mu$m determined by the SED fitting exceed 0.5.
	The definition of the quasar fraction is also provided in Section \ref{sec:SED_fit}.
	
\begin{figure}
		\centering
		\includegraphics[width=\columnwidth]{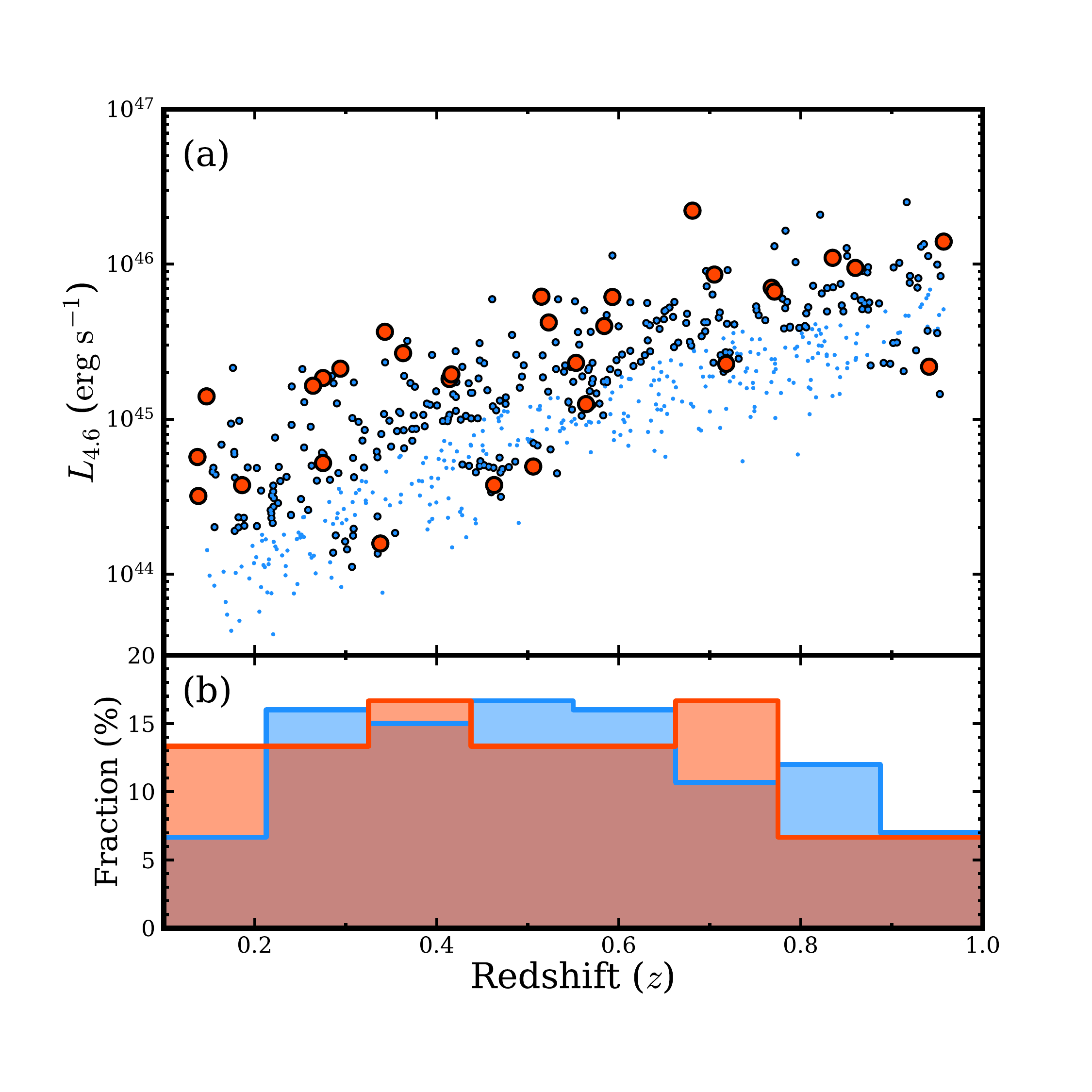}\\
		\caption{
			(a) Redshift versus $L_{\rm 4.6}$ luminosity. 
			Red circles and blue dots denote the selected dust-obscured quasars and SDSS quasars, respectively.
			Among the blue dots, 300 blue dots with black edges represent the $L_{\rm 4.6}$-$z$ matched SDSS quasars,
			which are used as unobscured quasars.
			(b) Redshift distributions of the two kinds of quasars.
			Red and blue histograms represent the dust-obscured quasars and the unobscured quasars, respectively.
			\label{fig:sample}}
	\end{figure}

	The selected 30 dust-obscured quasars span over wide ranges of BH mass ($10^{8.15}\,M_{\odot} < M_{\rm BH} < 10^{9.22}\,M_{\odot}$),
	bolometric luminosity ($10^{44.94}\,{\rm erg~s^{-1}} < L_{\rm bol} < 10^{47.07}\,{\rm erg~s^{-1}}$), and redshift ($0.137 < z < 0.957$),
	where the BH masses and bolometric luminosities are derived using $L_{\rm MIR}$-based-estimators \citep{kim10,kim23}.
	Their basic properties are shown in Figure \ref{fig:sample}.
	
	In order to compare the dust-obscured quasars to unobscured quasars,
	we select unobscured quasars from SDSS Data Release 14 (DR14) quasar catalog \citep{paris18}.
	This catalog contains 526,265 quasars, and their spectral properties were measured \citep{rakshit20}
	by performing multi-component fitting with \texttt{PyQSOFit} code \citep{guo19,shen19}.
	To avoid sample bias effects, we choose the SDSS quasars with the same selection criteria of the dust-obscured quasars: 
	(i) redshift range of $0.137 < z < 0.957$; 
	(ii) FIRST radio detection; 
	(iii) detections in the $J$-, $H$-, and $K$-bands from 2MASS PSC;
	and (iv) quasar fractions at 3.4\,$\mu$m and 4.6\,$\mu$m from the SED fitting exceed 0.5.
	
	Moreover, while SDSS quasars are generally known to have negligible dust extinction,
	\cite{kim23} and \cite{ykim24} showed that a significant fraction ($\sim$10--15\,\%) of SDSS quasars are affected by dust extinction. 
	Therefore, we perform the SED fitting (See Section \ref{sec:SED_fit}) on SDSS quasars 
	and include only those samples with the measured $E(B-V) < 0.1$.
	These SDSS quasars with $E(B-V) < 0.1$ can be classified as unobscured quasars.
	\cite{kim23} showed that the low $E(B-V)$ SDSS quasars exhibit similar Balmer decrement of unobscured quasars \citep{dong08},
	supporting the classification of these SDSS quasars with low $E(B-V)$ as unobscured.
	
	Using these selection criteria, 614 SDSS quasars are chosen. 
	Figure \ref{fig:sample} shows the $L_{\rm 4.6}$--$z$ distributions of the dust-obscured quasars and the selected SDSS quasars.
	The dust-obscured quasars tend to be slightly brighter than the SDSS quasars at a given redshift, which may cause a sample bias.
	To eliminate this effect, we select the 10 closest SDSS quasars to each dust-obscured quasar
	using a two-dimensional metric on the $L_{\rm 4.6}$ and redshift space.
	If an SDSS quasar was already selected for another nearby dust-obscured quasar, we excluded it and selected the next closest SDSS quasar.
	
	Finally, 300 SDSS quasars with similar $L_{\rm 4.6}$ values and redshifts to the selected 30 dust-obscured quasars are found
	and used as unobscured quasars.
	The unobscured quasars have a wide range of BH mass ($10^{7.76}\,M_{\odot} < M_{\rm BH} < 10^{10.03}\,M_{\odot}$),
	bolometric luminosity ($10^{44.79}\,{\rm erg~s^{-1}} < L_{\rm bol} < 10^{47.12}\,{\rm erg~s^{-1}}$), 
	and redshift ($0.154 < z < 0.954$).
	Naturally, their $K$-band magnitudes and redshift distributions are comparable to
	those of the dust-obscured quasars, as shown in Figure \ref{fig:sample}.
	
	\section{Analysis}\label{sec:analysis}
	\subsection{Spectral Fitting of Balmer Lines}\label{sec:Line_fit}
	
	In this subsection, we describe how H$\beta$ and H$\alpha$ lines of the dust-obscured quasars are fitted
	to measure FWHM values for broad components of the Balmer lines,
	and the FWHM values are used for estimating BH masses of the dust-obscured quasars.
	For this analysis, we use the optical and/or NIR spectra provided by \cite{glikman12}.
	However, note that the FWHM values of the unobscured quasars are adopted from \cite{rakshit20}.
	
	To fit the Balmer lines of the dust-obscured quasars, we use the \texttt{PyQSOFit} code
	for consistency in methodology, which is identical to the procedure for the unobscured quasars \citep{rakshit20}.
	Using the \texttt{PyQSOFit} code, 
	we decompose several components of host galaxy, power-law continuum, and \ion{Fe}{II} emissions from the spectra.
	After that, we fit the Balmer lines. 
	We fit H$\beta$ and H$\alpha$ line complexes separately,
	while all emission lines in each line complex are fitted simultaneously.
	Note that the H$\beta$ line complex includes H$\beta$ (broad and narrow components) 
	and [\ion{O}{III}] $\lambda \lambda$ 4959, 5007 doublet (core and wing components),
	and the H$\alpha$ line complex comprises H$\alpha$ (broad and narrow components), 
	[\ion{N}{II}] $\lambda \lambda$ 6549, 6585 doublet, and [\ion{S}{II}] $\lambda \lambda$ 6718, 6732 doublet.
	Although the flux ratios of these doublets are fixed,
	we fit the Balmer line complexes without fixing the flux ratios due to low S/N and partially truncated spectra.
	We find that not fixing the flux ratios has a negligible effects on
	the FWHM measurements of the broad components of the Balmer lines.
	
	During the fitting, the broad components of the Balmer lines are fitted with a single- or multiple-Gaussian model.
	However, the FWHM values estimated by the single-Gaussian model is systematically different from
	the FWHM values from the multi-Gaussian model \citep{kim10,shen11}.
	Thus, we apply correction factors \citep{kim10} to remove the model bias,
	which are derived from well-resolved Balmer lines of 26 unobscured bright quasars at low-$z$.
	For the broad components, we set a lower limit of the FWHM as 2000\,$\rm km~s^{-1}$,
	and this FWHM criterion for the broad components has been widely used (e.g., \citealt{suh19,xu07}).
	The FWHM criterion, broader than that used in \cite{glikman12}, 
	is set to exclude narrow line Seyfert 1 galaxies (e.g., $\rm FWHM \lesssim 2000\,km~s^{-1}$; \citealt{zhou06})
	that may have systematically different $\lambda_{\rm Edd}$ values \citep{boroson02}.
	Moreover, we only use the FWHM values that are measured reliably,
	selecting those with a $\rm FWHM$/$\rm FWHM_{fit\_err}$ of greater than 3,
	where $\rm FWHM_{fit\_err}$ represents the error in FWHM from the fitting. 
	
	Through this process, we fit the spectra of 120 dust-obscured quasars in \cite{glikman12},
	ultimately selecting 30 objects with reliably measured FWHM values.
	Figure \ref{fig:fit} shows the line fitting results.
	The fit provides the FWHM values of the broad components of the Balmer lines,
	but we use the FWHM values after correcting instrumental wavelength dispersion as
	$\rm FWHM^2 = (FWHM_{fit})^2 - (FWHM_{inst})^2$,
	where $\rm FWHM_{fit}$ and $\rm FWHM_{inst}$ denote 
	the FWHM provided from the fit and the wavelength dispersion due to the instrumental spectral resolution, respectively.
	In this work, we use the corrected FWHM values of the Balmer lines,
	and they are summarized in Table \ref{tbl:fit}.
	Note that to avoid the effects of correction bias, the same instrumental wavelength dispersion correction is also applied to
	the FWHM values of unobscured quasars adopted from \cite{rakshit20},
	and the corrected FWHM values are used in this study.
	
	\begin{figure*}
		\centering
		\includegraphics[width=\textwidth]{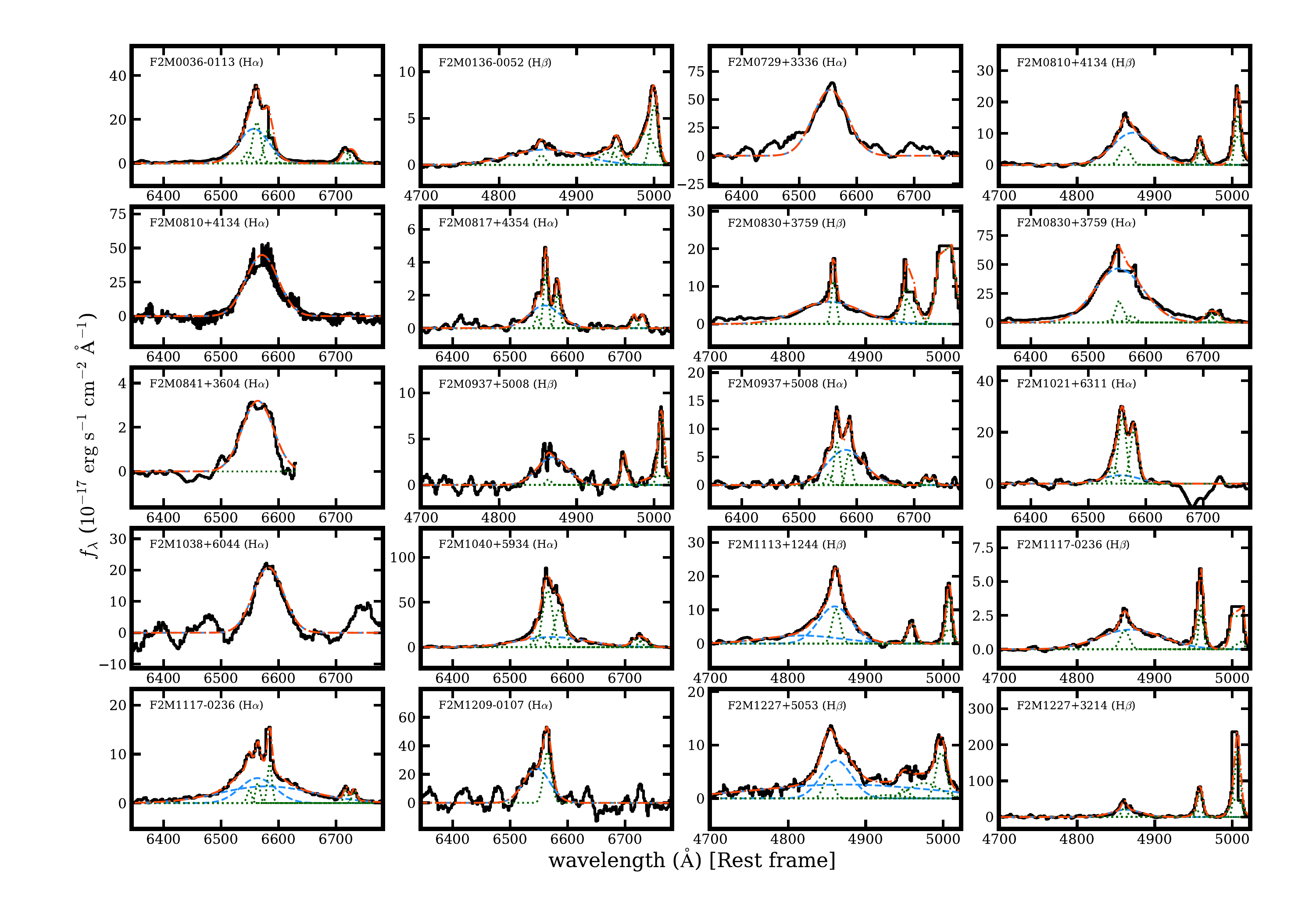}\\
		\caption{
			The fitting results of the H$\beta$ and H$\alpha$ lines for the dust-obscured quasars. 
			The black solid lines denote the continuum-subtracted spectra in the rest-frame. 
			The red dash-dotted lines are the best-fit model,
			and the blue dashed and green dotted lines represent the broad and narrow components of the each line complex, respectively.
			\label{fig:fit}}
	\end{figure*}
	
	\setcounter{figure}{2}
	\begin{figure*}
		\centering
		\includegraphics[width=\textwidth]{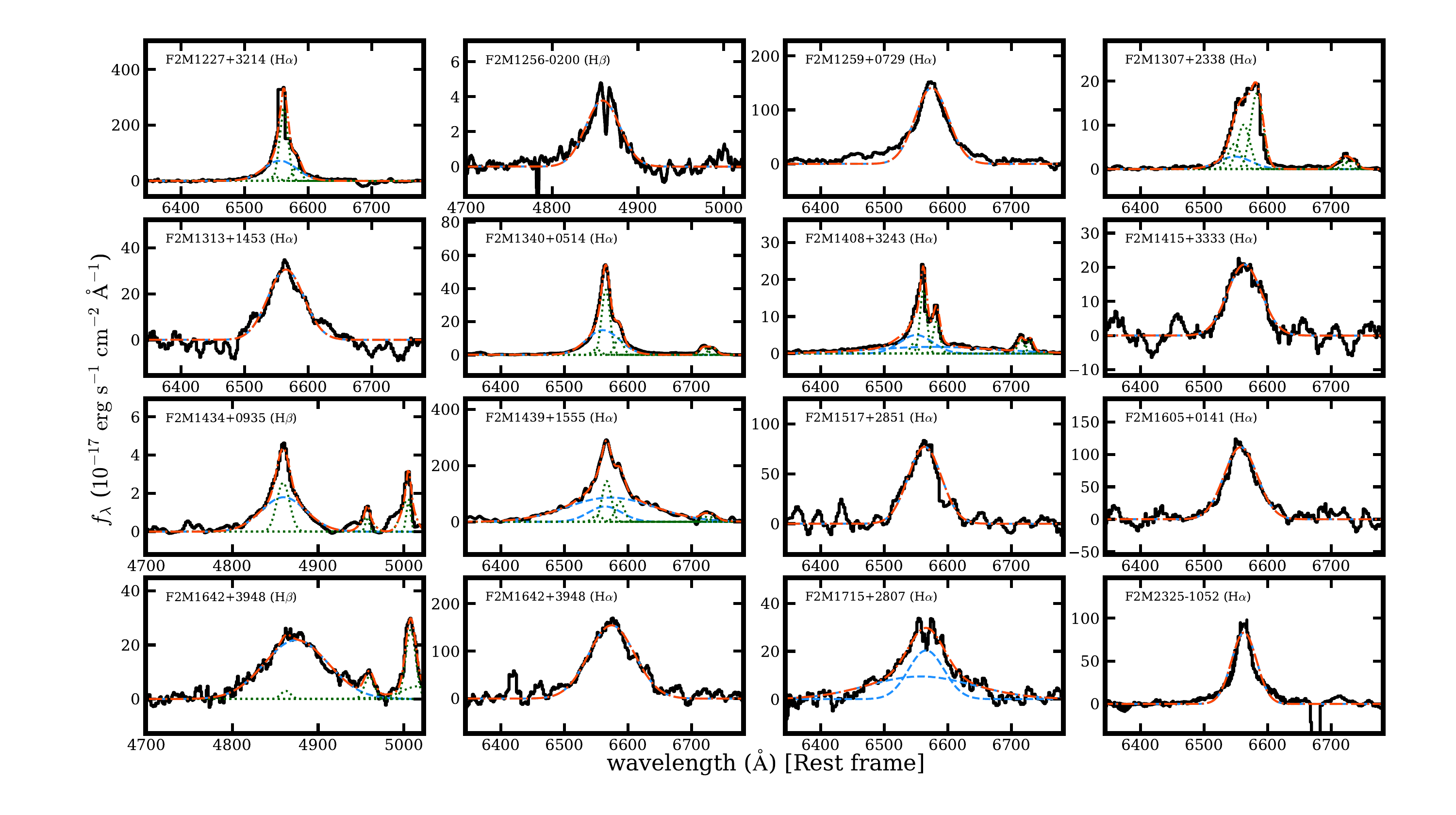}\\
		\caption{
			Continued.}
	\end{figure*}
	
\begin{figure*}
	\centering
	\includegraphics[width=0.8\textwidth]{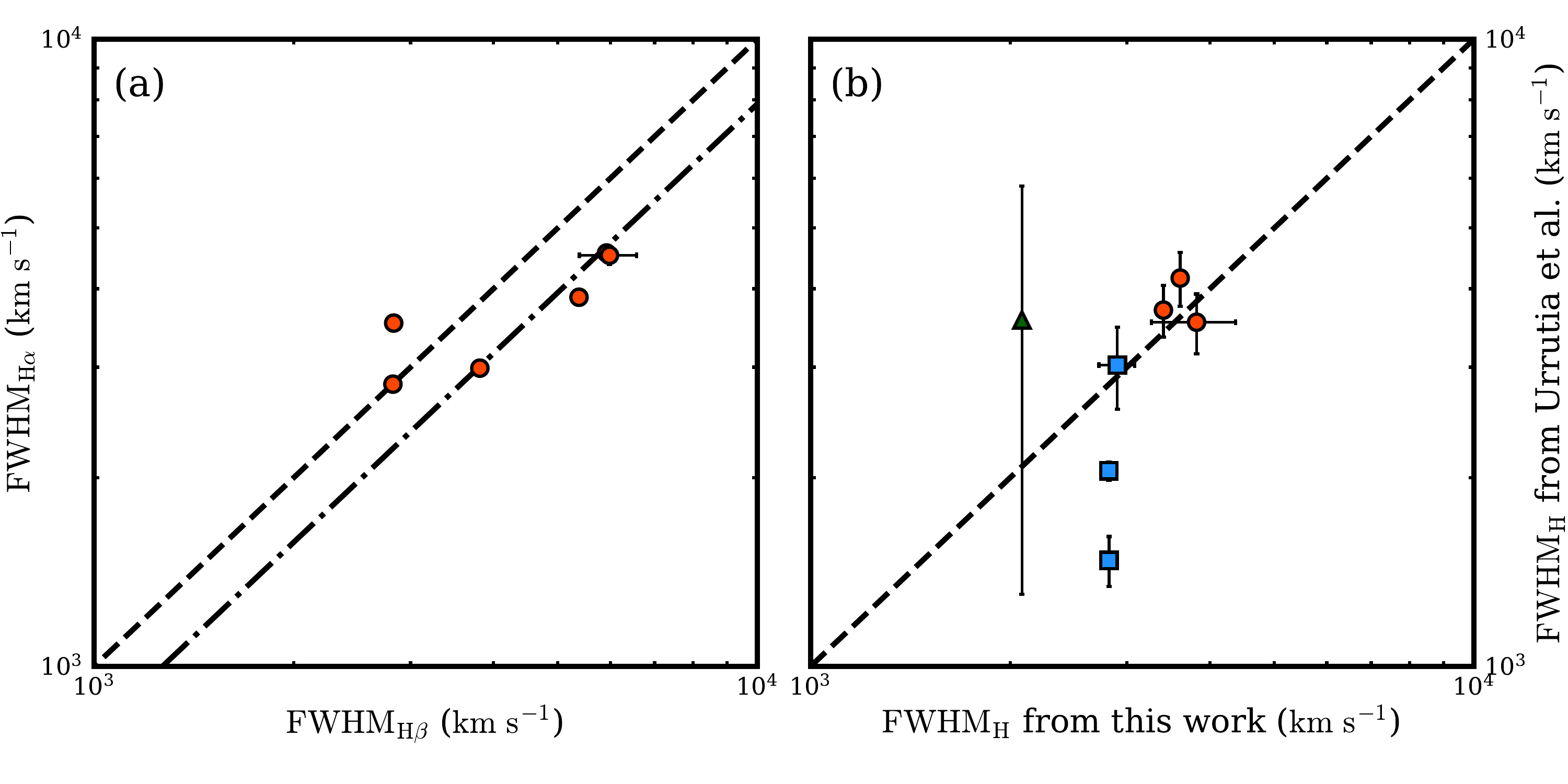}\\
	\caption{
		(a) Comparison of FWHM values of broad components of H$\beta$ and H$\alpha$ lines.
		The dashed line denotes a line where the FWHM values of the Balmer lines are identical,
		and dash-dotted line is a well-known relationship between the FWHM values of the Balmer lines \citep{bisogni17}.
		(b) Comparison of Balmer line FWHM values measured from this work and \cite{urrutia12}.
		Red circles denote the comparison of $\rm FWHM_{H\beta}$ measurements,
		and blue squares mean the comparison of $\rm FWHM_{H\alpha}$ measurements.
		Green triangles represent the $\rm FWHM_{H\alpha}$ measurement comparison,
		but the $\rm FWHM_{H\beta}$ measured in \cite{urrutia12} is converted to the $\rm FWHM_{H\alpha}$
		using the relationship established in \cite{bisogni17}.
		The meaning of dashed line is identical to that of the panel (a).
		\label{fig:FWHM_HbHa}}
	\end{figure*}

	We validate the accuracy of our FWHM measurements through two methods:
	(i) by comparing them with the well-established relationship between Balmer lines;
	(ii) by cross-referencing with individual measurements from previous studies.
	Among the 30 dust-obscured quasars, 
	the fit provides the FWHM values for both the H$\beta$ and H$\alpha$ lines in six objects.
	Our FWHM measurements for the Balmer lines are consistent with 
	the well-known relationship, $\log \left( \frac{\rm FWHM_{H\beta}}{\rm km~s^{-1}} \right) = 0.103 + \log \left( \frac{\rm FWHM_{H\alpha}}{\rm km~s^{-1}} \right)$, 
	established by \cite{bisogni17}, as shown in Figure \ref{fig:FWHM_HbHa}.
	Additionally, seven dust-obscured quasars were studied in \cite{urrutia12}, and their Balmer line FWHM values were measured.
	Among these objects, three have their $\rm FWHM_{H\beta}$ values measured in both this work and \cite{urrutia12},
	while another three objects have $\rm FWHM_{H\alpha}$ values measured in both studies.
	For the remaining object, $\rm FWHM_{H\alpha}$ is measured in this work, whereas \cite{urrutia12} measured $\rm FWHM_{H\beta}$.
	For this object, we convert the $\rm FWHM_{H\beta}$ to $\rm FWHM_{H\alpha}$ using the relationship established by \cite{bisogni17} for comparison.
	Comparing the FWHM values measured in both studies reveals that,
	although not perfectly aligned, they closely match for most objects, with only a few exceptions.
	Upon selecting only those samples whose both FWHM measurements are over 2000\,$\rm km~s^{-1}$,
	the rms scatter between the FWHM measurements is 0.12\,dex, which is taken as the average FWHM error ($\rm FWHM_{avg\_err}$).
	The total error in FWHM ($\rm FWHM_{tot\_err}$) is taken as the square root of the quadratic sum of the fitting error and average error
	as $\rm FWHM^2_{tot\_err} = FWHM^2_{fit\_err} + FWHM^2_{avg\_err}$.
	However, the measured rms scatter is significantly affected by
	a few samples with the large discrepancies in FWHM measurements,
	suggesting the possibility that the rms scatter could be overestimated.
	Note that the exceptions are related to the $\rm FWHM_{H\alpha}$ measurements,
	which implies that the possibility of inaccuracy in the $\rm FWHM_{H\alpha}$ measurements cannot be ruled out.
	These comparisons of the FWHM values are shown in Figure \ref{fig:FWHM_HbHa},
	suggesting the reliability of our FWHM measurements.
	
	\begin{table*}
	\centering
	\caption{Line and SED fitting results of dust-obscured quasars \label{tbl:fit}}
	\begin{tabular}{ccccccccc}
		\hline\hline
		\noalign{\smallskip}
		object name&                            redshift&                       &
		$\rm FWHM_{\rm H\beta}$\tablefootmark{$\dagger$}&       $\rm FWHM_{H\alpha}$\tablefootmark{$\dagger$}&    &
		$\log L_{\rm 3.4}$\tablefootmark{$\ast$}& 					$\log L_{\rm 4.6}$\tablefootmark{$\ast$}&		$E(B-V)$   \\
		&                                               	&                          &
		($\rm{km~s^{-1}}$)&            			($\rm km~s^{-1}$)&     &
		($\rm{erg~s^{-1}}$)&            		($\rm erg~s^{-1}$)&  	(mag.)\\
		\noalign{\smallskip}
		\hline
		\noalign{\smallskip}
		
		F2M0036$-$0113&		0.294&	&	--&							2818$\pm$860&		&45.34$\pm$0.02&		45.33$\pm$0.02&		1.60$\pm$0.01\\
		F2M0136$-$0052&		0.718&	&	6775$\pm$2109&	--&								&45.37$\pm$0.02&		45.36$\pm$0.02&		0.48$\pm$0.01\\
		F2M0729$+$3336&		0.957&	&	--&							3182$\pm$970&		&46.16$\pm$0.03&		46.15$\pm$0.03&		0.50$\pm$0.01\\
		F2M0810$+$4134&		0.506&	&	3815$\pm$1164&	2989$\pm$912&		&44.71$\pm$0.03&	44.69$\pm$0.03&		0.21$\pm$0.01\\
		F2M0817$+$4354&		0.186&	&	--&							2901$\pm$903&		&44.59$\pm$0.02&	44.57$\pm$0.02&		0.80$\pm$0.01\\
		F2M0830$+$3759&		0.414&	&	5923$\pm$1809&	4562$\pm$1392&		&45.28$\pm$0.02&	45.26$\pm$0.02&		0.61$\pm$0.01\\
		F2M0841$+$3604&		0.553&	&	--&							2902$\pm$902&		&45.38$\pm$0.02&	45.36$\pm$0.02&		0.79$\pm$0.01\\
		F2M0937$+$5008&		0.275&	&	2829$\pm$863&	3527$\pm$1076&		&44.73$\pm$0.02&	44.72$\pm$0.02&		0.46$\pm$0.01\\
		F2M1021$+$6311&		0.138&	&	--&							3521$\pm$1074&		&44.52$\pm$0.02&		44.50$\pm$0.02&		0.65$\pm$0.01\\
		F2M1038$+$6044&		0.515&	&	--&							2804$\pm$855&		&45.81$\pm$0.02&		45.79$\pm$0.02&		0.87$\pm$0.01\\
		F2M1040$+$5934&		0.147&	&	--&							6468$\pm$1973&		&45.16$\pm$0.02&		45.15$\pm$0.02&		0.95$\pm$0.01\\
		F2M1113$+$1244&		0.681&	&	3404$\pm$1038&	--&								&46.36$\pm$0.02&		46.35$\pm$0.02&		0.90$\pm$0.01\\
		F2M1117$-$0236&		0.463&	&	5983$\pm$1918&	4523$\pm$1387&		&44.59$\pm$0.04&		44.57$\pm$0.04&		0.21$\pm$0.01\\
		F2M1209$-$0107&		0.363&	&	--&							2394$\pm$730&			&45.44$\pm$0.03&		45.43$\pm$0.03&		0.87$\pm$0.01\\
		F2M1227$+$5053&		0.768&	&	3608$\pm$1101&	--&								&45.86$\pm$0.03&		45.85$\pm$0.03&		0.33$\pm$0.01\\
		F2M1227$+$3214&		0.137&	&	2822$\pm$861&	2818$\pm$860&		&44.77$\pm$0.02&		44.76$\pm$0.02&		0.79$\pm$0.01\\
		F2M1256$-$0200&		0.835&	&	2900$\pm$885&	--&								&46.06$\pm$0.02&		46.04$\pm$0.02&		0.80$\pm$0.01\\
		F2M1259$+$0729&		0.860&	&	--&							2802$\pm$855&		&45.99$\pm$0.02&		45.98$\pm$0.02&		0.37$\pm$0.01\\
		F2M1307$+$2338&		0.275&	&	--&							2816$\pm$859&		&45.28$\pm$0.02&		45.27$\pm$0.02&		1.17$\pm$0.01\\
		F2M1313$+$1453&		0.584&	&	--&							3028$\pm$924&		&45.62$\pm$0.02&		45.60$\pm$0.02&		0.62$\pm$0.01\\
		F2M1340$+$0514&		0.264&	&	--&							2820$\pm$860&		&45.23$\pm$0.02&		45.22$\pm$0.02&		0.95$\pm$0.01\\
		F2M1408$+$3243&		0.338&	&	--&							3481$\pm$1062&		&44.21$\pm$0.04&		44.20$\pm$0.04&		0.42$\pm$0.01\\
		F2M1415$+$3333&		0.416&	&	--&							2854$\pm$870&		&45.31$\pm$0.02&		45.29$\pm$0.02&		0.61$\pm$0.01\\
		F2M1434$+$0935&		0.771&	&	3819$\pm$1289&	--&								&45.84$\pm$0.02&		45.82$\pm$0.02&		0.68$\pm$0.01\\
		F2M1439$+$1555&		0.941&	&	--&							4960$\pm$1513&		&45.36$\pm$0.04&		45.34$\pm$0.04&		0.23$\pm$0.02\\
		F2M1517$+$2851&		0.705&	&	--&							2798$\pm$853&		&45.95$\pm$0.02&		45.93$\pm$0.02&		0.64$\pm$0.01\\
		F2M1605$+$0141&		0.343&	&	--&							2796$\pm$853&		&45.58$\pm$0.02&		45.56$\pm$0.02&		1.02$\pm$0.01\\
		F2M1642$+$3948&		0.593&	&	5383$\pm$1642&	3878$\pm$1183&		&45.81$\pm$0.03&		45.79$\pm$0.03&		0.16$\pm$0.01\\
		F2M1715$+$2807&		0.523&	&	--&							3828$\pm$1168&		&45.64$\pm$0.02&		45.62$\pm$0.02&		0.91$\pm$0.01\\
		F2M2325$-$1052&		0.564&	&	--&							2083$\pm$635&			&45.11$\pm$0.03&		45.10$\pm$0.03&		0.38$\pm$0.01\\
		
		\noalign{\smallskip}
		\hline
	\end{tabular}
	\tablefoot{\\
	\tablefoottext{$\dagger$}{The errors of $\rm FWHM_{H\beta}$ and $\rm FWHM_{H\alpha}$ are their $\rm FWHM_{tot\_err}$ values.}\\
	\tablefoottext{$\ast$}{Dust extinction corrected luminosities.}
	}
	\end{table*}
	
	\subsection{SED Fitting for MIR Continuum Luminosities}\label{sec:SED_fit}
	In order to measure $L_{\rm bol}$ and $M_{\rm BH}$ values of the dust-obscured and the unobscured quasars
	using the $L_{\rm MIR}$-based estimators \citep{kim23},
	we estimate monochromatic continuum luminosity, $\lambda L_{\lambda}$, 
	at 3.4\,$\mu$m and 4.6\,$\mu$m (hereafter, $L_{\rm 3.4}$ and $L_{\rm 4.6}$, respectively) in the rest-frame.
	Although \cite{kim23} showed that host galaxy contaminations at 3.4\,$\mu$m and 4.6\,$\mu$m are negligible ($<20\,\%$),
	we measure the $L_{\rm 3.4}$ and $L_{\rm 4.6}$ values after subtracting the host galaxy light by an SED fitting
	with optical-to-MIR photometric data of SDSS, 2MASS PSC, and WISE.
	For these photometric data, the intrinsic fluctuation ($\sigma_{m}=0.035$\,mag) 
	of the spectrum with respect to a simple power-law spectrum found in \cite{kim15b}
	is added in quadrature to the original photometric error in each band.
	Note that we additionally use $B$- and $R$-band photometry \citep{lasker08} adopted from \cite{glikman12}
	for only the dust-obscured quasars.
	
	The SED fitting is implemented using the method described in \cite{kim23}.
	The photometric data, denoted as $f(\lambda)$, are fitted with an SED model
	that combines quasar ($Q(\lambda)$), elliptical galaxy ($E(\lambda)$), spiral galaxy ($S(\lambda)$), and irregular galaxy ($I(\lambda)$)
	spectra through a weighted sum, as
	\begin{equation}
		f(\lambda) = C_{1}Q(\lambda) + C_{2}E(\lambda) + C_{3}S(\lambda) + C_{4}I(\lambda), 
	\end{equation}
	where $C_{1}$, $C_{2}$, $C_{3}$, and $C_{4}$ are the normalization constants.
	Note that the $Q(\lambda)$, $E(\lambda)$, $S(\lambda)$, and $I(\lambda)$
	are reddened spectra of $Q_{0}(\lambda)$, $E_{0}(\lambda)$, $S_{0}(\lambda)$, and $I_{0}(\lambda)$, respectively.
	The $Q_{0}(\lambda)$ spectrum is adopted from \cite{krawczyk13},
	while the $E_{0}(\lambda)$, $S_{0}(\lambda)$, and $I_{0}(\lambda)$ spectra are taken from \cite{assef10}.
	The intrinsic quasar spectrum of \cite{krawczyk13} is not significantly different from 
	other quasar spectral templates (e.g., \citealt{richards06,assef10}) in the optical-to-MIR wavelength range \citep{kim23}.
	
	For the dust reddening, we adopt a reddening law of \cite{fitzpatrick99}
	based on average Galactic extinction curve with an assumption of $R_V = 3.1$ (e.g., \citealt{weingartner01}).
	Note that the reddening law is significantly different from other reddening laws (e.g., \citealt{calzetti00}) in UV region,
	particularly due to the presence/absence of UV bump at 2175\,$\rm \AA{}$,
	which can alter the SED fitting results depending on the reddening law used.
	Therefore, the UV region ($< 4000\,{\rm \AA{}}$) is excluded from our SED fitting.
	
\begin{figure*}
	\centering
	\includegraphics[width=\textwidth]{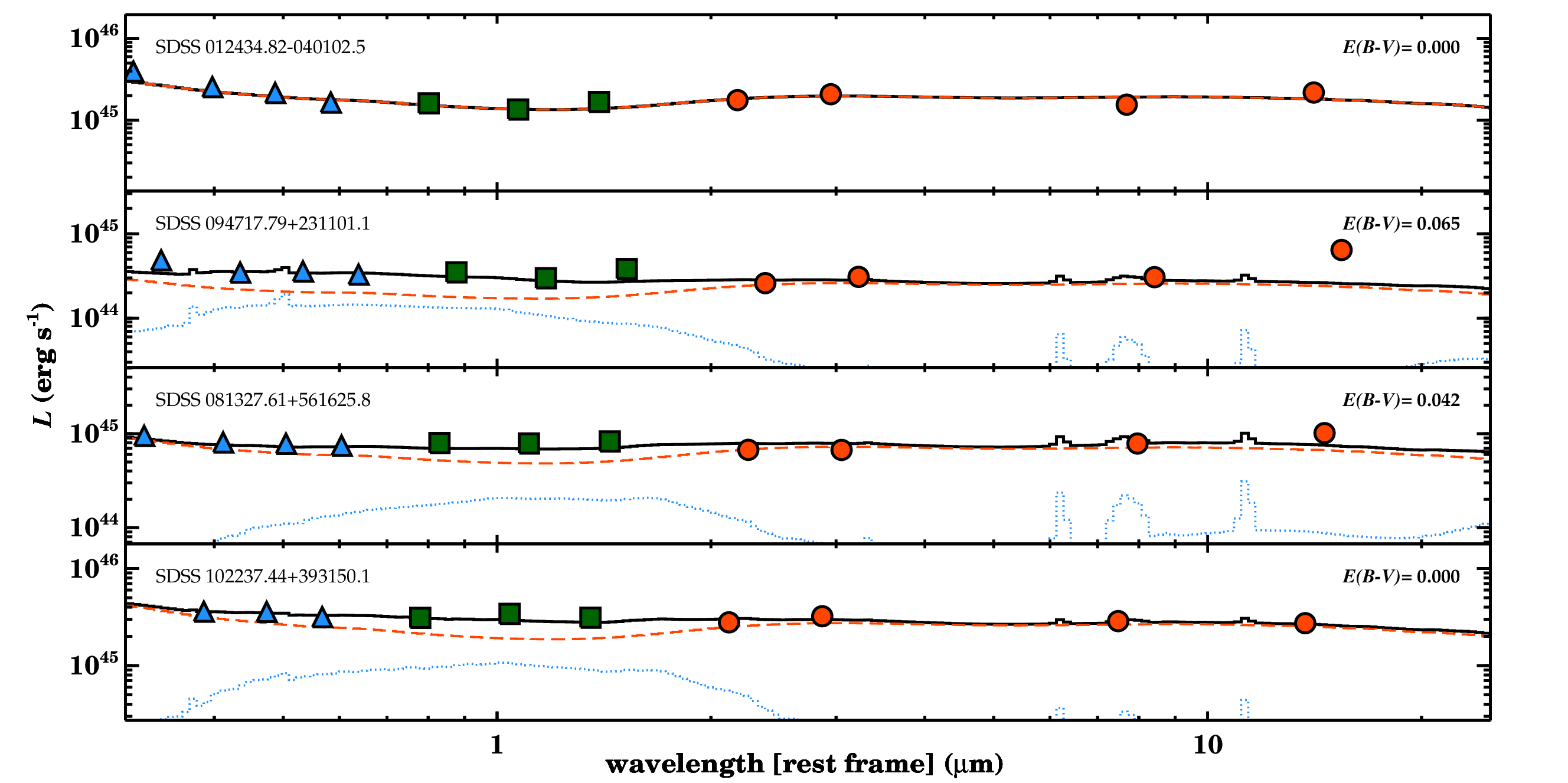}\\
	\caption{
		Best-fit SED models and photometric data of four randomly selected unobscured quasars.
		Blue triangles, green squares, and red circles denote the photometric data points of SDSS, 2MASS PSC, and WISE, respectively.
		Red dashed and blue dotted lines mean the reddened spectra of quasar and host galaxy, respectively,
		and black solid lines represent the best-fit SED models.
		The name of the randomly selected sample and their $E(B-V)$ values measured from the SED fitting
		are presented on the top-left and -right side of each panel, respectively.
		\label{fig:SED_UQ}}
	\end{figure*}

\begin{figure*}
	\centering
	\includegraphics[width=\textwidth]{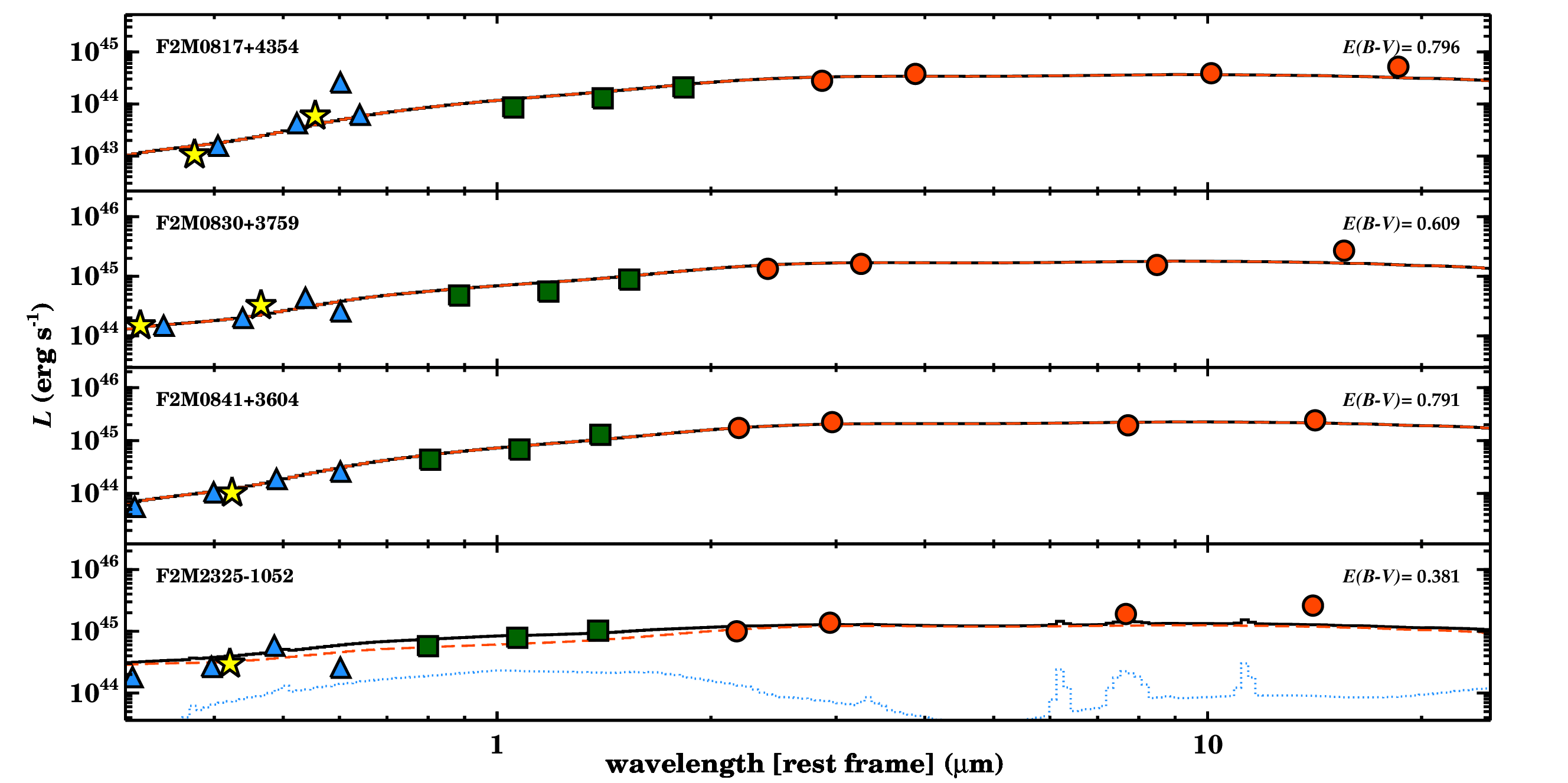}\\
	\caption{
		Best-fit SED models and photometric data of four randomly selected dust-obscured quasars.
		The meanings of the symbols and lines are identical to those in Figure \ref{fig:SED_UQ},
		but $B$- and $R$-band photometry are additionally shown as yellow stars.
		\label{fig:SED_DQ}}
	\end{figure*}

	For the SED fitting, an interactive data language (IDL) procedure, \texttt{MPFIT} \citep{markwardt09}, is used.
	Figure \ref{fig:SED_UQ} and \ref{fig:SED_DQ} show examples of the best-fit SED models 
	with the photometric data of unobscured and dust-obscured quasars, respectively.
	Through the SED fitting, we obtain the $E(B-V)$ values and extinction-corrected $L_{\rm 3.4}$ and $L_{\rm 4.6}$ values,
	and these properties of the dust-obscured quasars are presented in Table \ref{tbl:fit}.
	Moreover, since \cite{kim23} found that the measured $L_{\rm 3.4}$ and $L_{\rm 4.6}$ values based on the quasar spectrum of \cite{krawczyk13}
	are $\sim$5\,\% higher than those from \cite{assef10},
	the discrepancy is added as an uncertainty in quadrature to their original uncertainties obtained from the fit.
	Furthermore, through the SED fitting, we can measure quasar fractions, defined as $C_{1}Q(\lambda) / f(\lambda)$, at 3.4\,$\mu$m and 4.6\,$\mu$m.
	Since quasar activities of the dust-obscured and the unobscured quasars are dominant,
	we only use the SED fitting results for the sample with the measured quasar fractions exceeding 0.5.

	\subsection{$L_{\rm bol}$ and $M_{\rm BH}$ values}\label{sec:prop}
	In this subsection, we estimate bolometric luminosities and black hole masses for both the dust-obscured and the unobscured quasars.
	However, given that dust-obscured quasars are heavily obscured,
	it is necessary to use $L_{\rm bol}$ and $M_{\rm BH}$ estimators that are immune to the dust extinction.
	Therefore, we use the $L_{\rm MIR}$-based $L_{\rm bol}$ and $M_{\rm BH}$ estimators \citep{kim23}
	that can measure these properties without the effects of the dust extinction.
	
	For estimating the $L_{\rm bol}$ values,
	we use the extinction-corrected $L_{\rm 4.6}$ values measured from the SED fitting
	with the $L_{\rm 4.6}$-based $L_{\rm bol}$ estimator \citep{kim23}.
	The used $L_{\rm 4.6}$-based $L_{\rm bol}$ estimator is
	\begin{equation}
		\begin{aligned}
			\log \left( \frac{L_{\rm bol}}{\rm 10^{44}\,erg~s^{-1}} \right) = & (0.739\pm0.021) \\
		  	& + (0.993\pm0.031) \log \left( \frac{L_{\rm 4.6}}{\rm 10^{44}\,erg~s^{-1}} \right). \label{eqn:Lbol_46}
		\end{aligned}
	\end{equation}
	The measured $L_{\rm bol}$ values of the dust-obscured quasars are listed in Table \ref{tbl:prop}.
	
	To obtain the $M_{\rm BH}$ values,
	we use the $L_{\rm MIR}$-based $M_{\rm BH}$ estimators \citep{kim23}, which are
	\begin{equation}
		\begin{aligned}
			\log \left( \frac{M_{\rm BH}}{M_{\odot}} \right) = & \left( 6.914\pm0.023 \right) + 
			\left( 0.478\pm0.016 \right) \log \left( \frac{L_{\rm 4.6}}{\rm 10^{44}\,erg~s^{-1}} \right)\\
			& + 2 \log \left( \frac{\rm FWHM_{H\beta}}{\rm 1000\,km~s^{-1}} \right)		\label{eqn:MBH_46_Hb}		
		\end{aligned}
	\end{equation}
	and
	\begin{equation}
		\begin{aligned}
			\log \left( \frac{M_{\rm BH}}{M_{\odot}} \right) = & \left( 6.973\pm0.060 \right) + 
			\left( 0.478\pm0.016 \right) \log \left( \frac{L_{\rm 4.6}}{\rm 10^{44}\,erg~s^{-1}} \right)\\
			& + \left( 2.06\pm0.06 \right) \log \left( \frac{\rm FWHM_{H\alpha}}{\rm 1000\,km~s^{-1}} \right).	\label{eqn:MBH_46_Ha}
		\end{aligned}
	\end{equation}
	For objects where both $\rm FWHM_{H\beta}$ and $\rm FWHM_{H\alpha}$ measurements are available,
	we use the $M_{\rm BH}$ values derived from $\rm FWHM_{H\alpha}$.
	This is because, in the case of the dust-obscured quasars, 
	some objects exhibit a broad component in the H$\alpha$ line,
	while the broad component of the H$\beta$ line is absent,
	suggesting that the $\rm FWHM_{H\beta}$ could be affected by dust extinction in extreme cases.
	The measured $M_{\rm BH}$ values of the dust-obscured quasars are summarized in Table \ref{tbl:prop}.
	
	Similar to the dust-obscured quasars, 
	we estimate the $M_{\rm BH}$ values of the unobscured quasars using Equations of \ref{eqn:MBH_46_Hb} and \ref{eqn:MBH_46_Ha}. 
	Among the 614 unobscured quasars selected in Section \ref{sec:sample}, 
	93 lack Balmer line measurements but exhibit \ion{Mg}{II} line properties.
	We convert their $\rm FWHM_{\ion{Mg}{II}}$ to $\rm FWHM_{H\beta}$ using the relationship presented in \cite{bisogni17},
	which are used to measure their $M_{\rm BH}$ values.
	
	
	\begin{table*}
		\centering
		\caption{BH properties of dust-obscured quasars \label{tbl:prop}}
		\begin{tabular}{ccccc}
			\hline\hline
			\noalign{\smallskip}
			object name&	&	{$\log \left( L_{\rm bol} / {\rm erg~s^{-1}} \right)$}& {$\log \left( M_{\rm BH} / M_{\odot}\right)$}& $\log \left( {\lambda_{\rm Edd}} \right)$ \\
			\noalign{\smallskip}
			\hline
			\noalign{\smallskip}
			
			F2M0036$-$0113&		&	46.06$\pm$0.06&		8.53$\pm$0.23&	$-$0.58$\pm${0.24}	\\
			F2M0136$-$0052&		&	46.09$\pm$0.06&		9.22$\pm${0.22}&	$-$1.24$\pm${0.22}	\\
			F2M0729$+$3336&		&	46.87$\pm$0.09&		9.03$\pm${0.24}&	$-$0.26$\pm${0.25}	\\
			F2M0810$+$4134&		&	45.43$\pm$0.05&		8.28$\pm${0.24}&	$-$0.96$\pm${0.24}	\\
			F2M0817$+$4354&		&	45.31$\pm$0.04&		8.20$\pm${0.23}&	$-$0.99$\pm${0.24}	\\
			F2M0830$+$3759&		&	45.99$\pm$0.06&		8.93$\pm${0.24}&	$-$1.04$\pm${0.24}	\\
			F2M0841$+$3604&		&	46.09$\pm$0.06&		8.58$\pm${0.24}&	$-$0.59$\pm${0.24}	\\
			F2M0937$+$5008&		&	45.45$\pm$0.04&		8.44$\pm${0.23}&	$-$1.09$\pm${0.24}	\\
			F2M1021$+$6311&		&	45.24$\pm$0.04&		8.34$\pm${0.23}&	$-$1.20$\pm${0.23}	\\
			F2M1038$+$6044&		&	46.52$\pm$0.08&		8.75$\pm${0.24}&	$-$0.33$\pm${0.24}	\\
			F2M1040$+$5934&		&	45.88$\pm$0.06&		9.19$\pm${0.24}&	$-$1.41$\pm${0.24}	\\
			F2M1113$+$1244&		&	47.07$\pm$0.09&		9.10$\pm${0.22}&	$-$0.13$\pm${0.23}	\\
			F2M1117$-$0236&		&	45.31$\pm$0.05&		8.60$\pm${0.24}&	$-$1.39$\pm${0.24}	\\
			F2M1209$-$0107&		&	46.15$\pm$0.07&		8.44$\pm${0.23}&	$-$0.38$\pm${0.24}	\\
			F2M1227$+$5053&		&	46.57$\pm$0.08&		8.91$\pm${0.22}&	$-$0.44$\pm${0.23}	\\
			F2M1227$+$3214&		&	45.49$\pm$0.04&		8.26$\pm${0.23}&	$-$0.87$\pm${0.23}	\\
			F2M1256$-$0200&		&	46.77$\pm$0.08&		8.81$\pm${0.22}&	$-$0.15$\pm${0.23}	\\
			F2M1259$+$0729&		&	46.70$\pm$0.08&		8.84$\pm${0.24}&	$-$0.24$\pm${0.24}	\\
			F2M1307$+$2338&		&	46.00$\pm$0.06&		8.50$\pm${0.23}&	$-$0.61$\pm${0.24}	\\
			F2M1313$+$1453&		&	46.33$\pm$0.07&		8.73$\pm${0.24}&	$-$0.50$\pm${0.24}	\\
			F2M1340$+$0514&		&	45.95$\pm$0.06&		8.48$\pm${0.23}&	$-$0.64$\pm${0.24}	\\
			F2M1408$+$3243&		&	44.94$\pm$0.04&		8.18$\pm${0.24}&	$-$1.35$\pm${0.24}	\\
			F2M1415$+$3333&		&	46.02$\pm$0.06&		8.53$\pm${0.23}&	$-$0.61$\pm${0.24}	\\
			F2M1434$+$0935&		&	46.55$\pm$0.08&		8.95$\pm${0.23}&	$-$0.50$\pm${0.24}	\\
			F2M1439$+$1555&		&	46.07$\pm$0.07&		9.05$\pm${0.24}&	$-$1.08$\pm${0.25}	\\
			F2M1517$+$2851&		&	46.66$\pm$0.08&		8.82$\pm${0.24}&	$-$0.26$\pm${0.24}	\\
			F2M1605$+$0141&		&	46.29$\pm$0.07&		8.64$\pm${0.24}&	$-$0.45$\pm${0.24}	\\
			F2M1642$+$3948&		&	46.51$\pm$0.08&		9.04$\pm${0.24}&	$-$0.63$\pm${0.25}	\\
			F2M1715$+$2807&		&	46.35$\pm$0.07&		8.95$\pm${0.24}&	$-$0.70$\pm${0.24}	\\
			F2M2325$-$1052&		&	45.83$\pm$0.06&		8.15$\pm${0.23}&	$-$0.43$\pm${0.24}	\\
			\noalign{\smallskip}
			\hline
		\end{tabular}
	\end{table*}
	
	\section{Eddington Ratios}\label{sec:Edd}
	Using the $L_{\rm bol}$ and $M_{\rm BH}$ values,
	we measure $\lambda_{\rm Edd}$ ($L_{\rm bol}$/$L_{\rm Edd}$, where $L_{\rm Edd}$ is Eddington luminosity) values
	of the dust-obscured and unobscured quasars.
	The measured $\lambda_{\rm Edd}$ values of the dust-obscured quasars are also listed in Table \ref{tbl:prop}.
	
\begin{figure}
		\centering
		\includegraphics[width=\columnwidth]{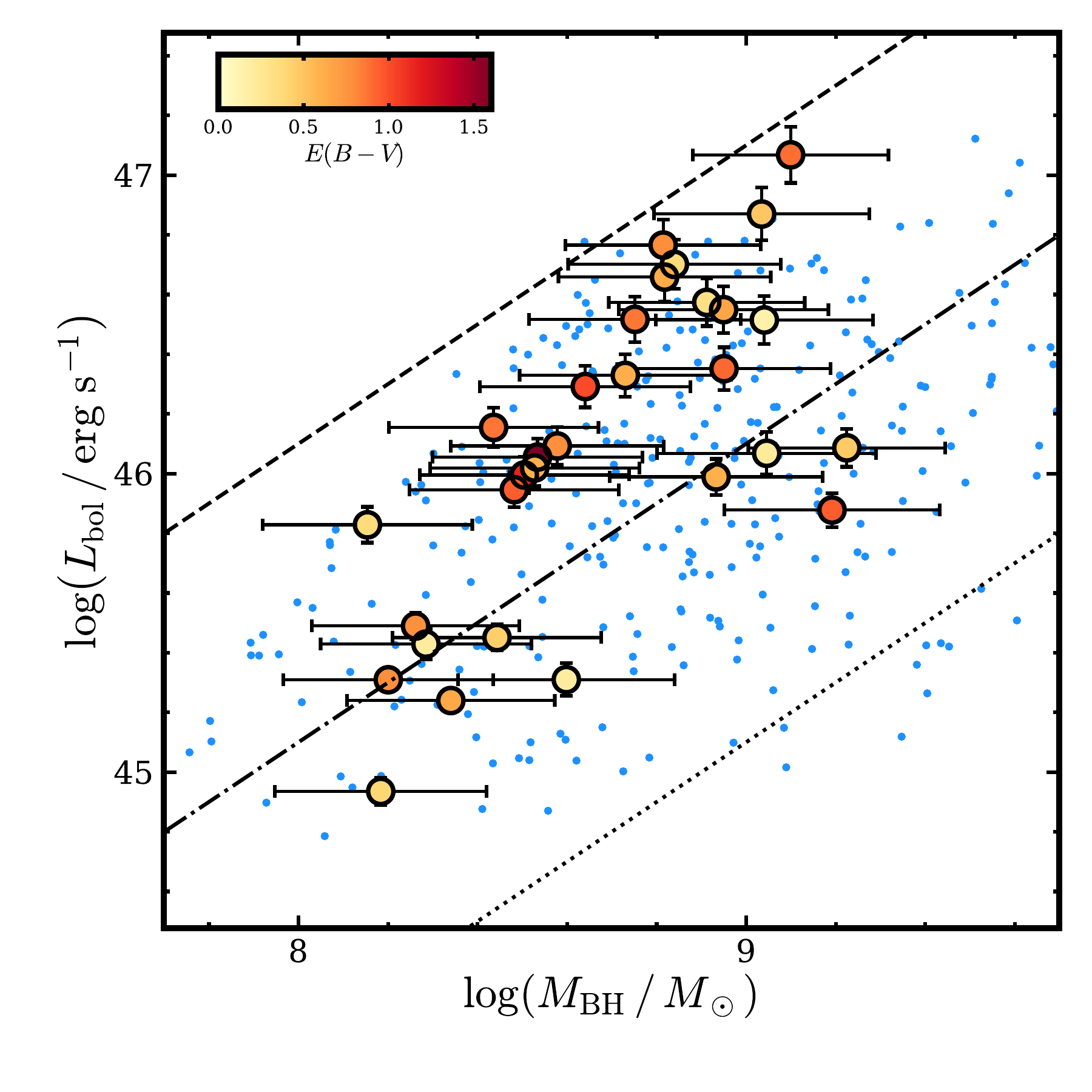}\\
		\caption{
			Bolometric luminosities versus BH masses of dust-obscured and unobscured quasars.
			The dust-obscured quasars are open circles with the colors representing their $E(B-V)$ values,
			and the unobscured quasars are blue dots.
			The dashed, dash-dotted, and dotted lines denote $\lambda_{\rm Edd}$ values of 1.0, 0.1, and 0.01, respectively.
			\label{fig:Edd1}}
	\end{figure}

\begin{figure}
		\centering
		\includegraphics[width=\columnwidth]{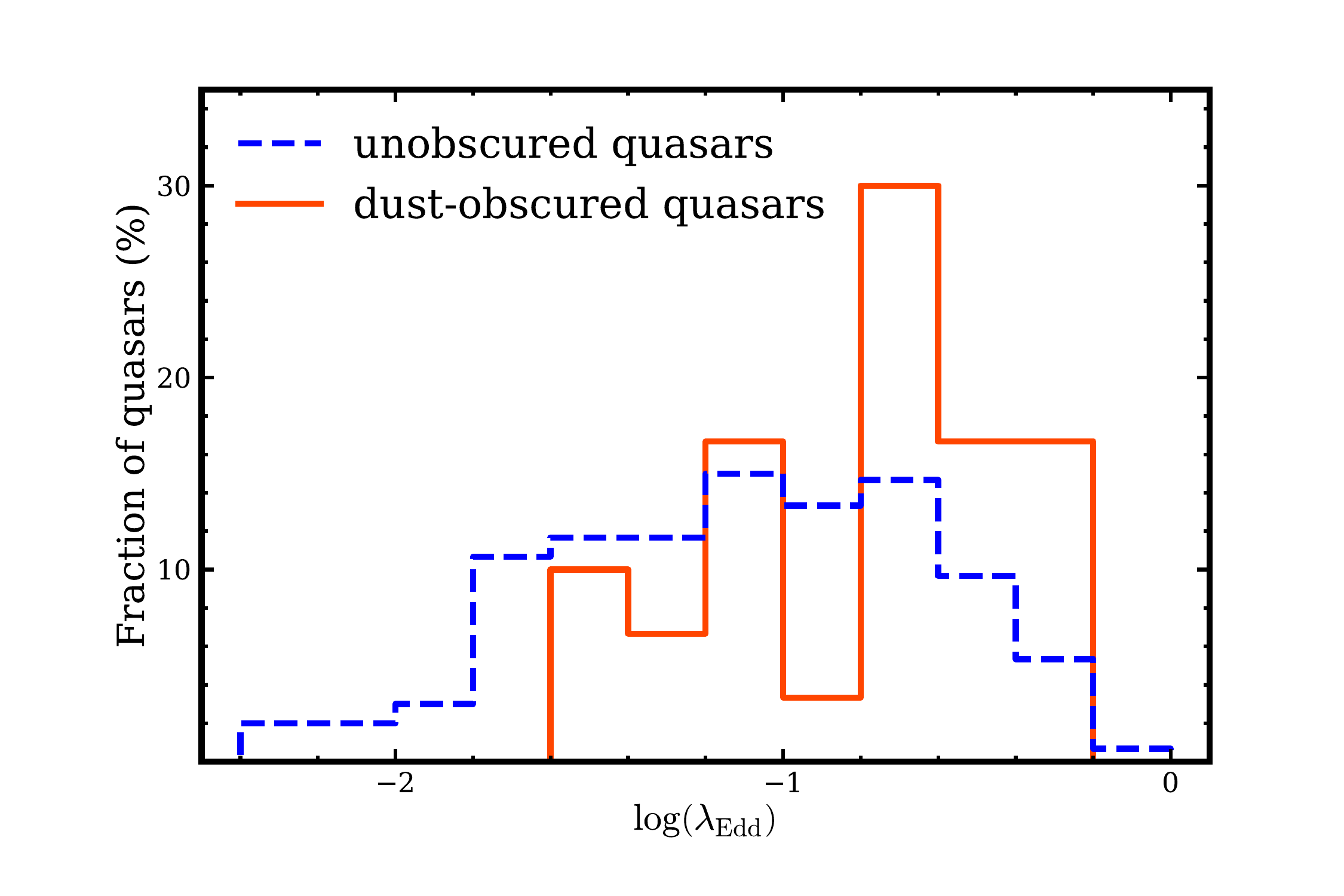}\\
		\caption{
			Distribution of $\lambda_{\rm Edd}$ values of dust-obscured and unobscured quasars.
			Red solid and blue dashed histograms mean the dust-obscured and the unobscured quasars, respectively.
			\label{fig:Edd2}}
	\end{figure}

	A comparison of the $L_{\rm bol}$ and $M_{\rm BH}$ values 
	of the dust-obscured and the unobscured quasars are shown in Figure \ref{fig:Edd1},
	and Figure \ref{fig:Edd2} shows distributions of the $\lambda_{\rm Edd}$ values.
	The median $\log \left( \lambda_{\rm Edd} \right)$ values of the dust-obscured and the unobscured quasars are
	$-$0.61$\pm$0.38 and $-$0.98$\pm$0.49, respectively,
	which shows that the $\lambda_{\rm Edd}$ values of the dust-obscured quasars are significantly higher than 
	those of the unobscured quasars. 
	In order to quantify how these $\lambda_{\rm Edd}$ values are significantly different,
	we perform a Kolmogorov-Smirnov test (K-S test) using \texttt{SciPy} package \citep{virtanen20} on Python.
	The maximum deviation between the cumulative distribution of these two $\lambda_{\rm Edd}$ values,
	$D$, is 0.36, and the probability of the result given the null hypothesis, $p$, is 0.001,
	indicating that they have different distributions.
	
\begin{figure*}
		\centering
		\includegraphics[width=0.8\textwidth]{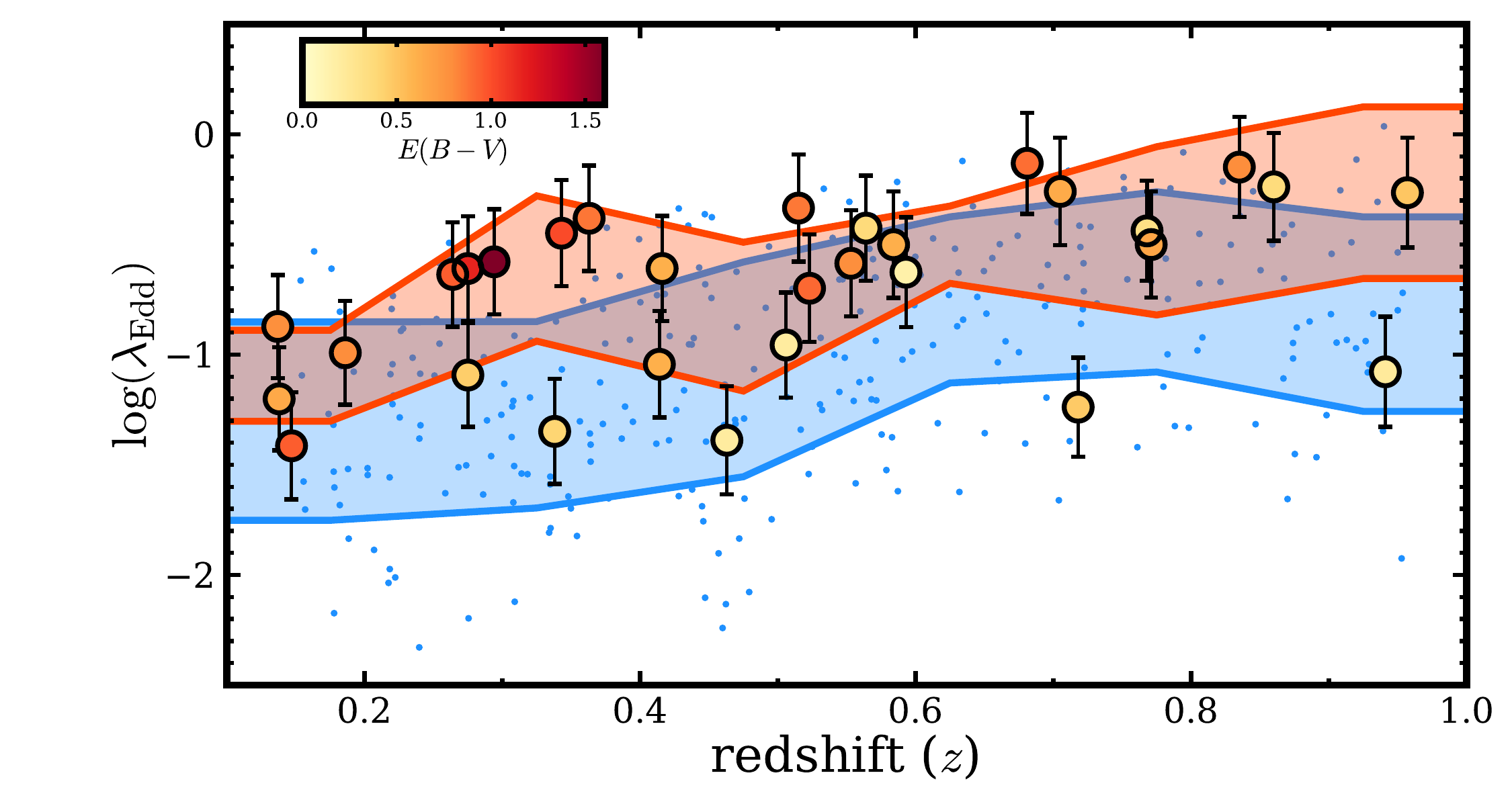}\\
		\caption{
			$\log \lambda_{\rm Edd}$ values vs. redshifts.
			The meanings of open circles and blue dots are identical to Figure \ref{fig:Edd1}.
			Red and blue polygons represent the $\log \lambda_{\rm Edd}$ values and their standard deviations
			of dust-obscured and unobscured quasars, respectively,
			which are divided into six sub-samples by redshifts. 
			\label{fig:Edd-z}}
	\end{figure*}

	Since $\lambda_{\rm Edd}$ values can vary with redshift
	due to sample bias effects or cosmological evolution,
	we compare the $\lambda_{\rm Edd}$ values of the two kinds of quasars across their redshifts.
	The comparison is shown in Figure \ref{fig:Edd-z}.
	For the comparison, we divide the two kinds of quasars into six bins of redshift:
	$0.10\sim0.25$, $0.25\sim0.40$, $0.40\sim0.55$, $0.55\sim0.70$, $0.70\sim0.85$, and $0.85\sim1.00$.
	The $\log \left( \lambda_{\rm Edd} \right)$ values of the dust-obscured quasars in the six redshift bins are
	$-$1.10$\pm$0.21, $-$0.61$\pm$0.33, $-$0.83$\pm$0.34, $-$0.50$\pm$0.18, $-$0.44$\pm$0.38, and $-$0.26$\pm$0.39,
	and those of the unobscured quasars are
	$-$1.30$\pm$0.45, $-$1.27$\pm$0.42, $-$1.07$\pm$0.49, $-$0.75$\pm$0.38, $-$0.67$\pm$0.41, and $-$0.82$\pm$0.44.
	This comparison reveals that
	the $\lambda_{\rm Edd}$ values of the dust-obscured quasars are significantly higher than
	those of the unobscured quasars for a wide redshift range ($0 \lesssim z \lesssim 1$).
	
\begin{figure*}
		\centering
		\includegraphics[width=0.8\textwidth]{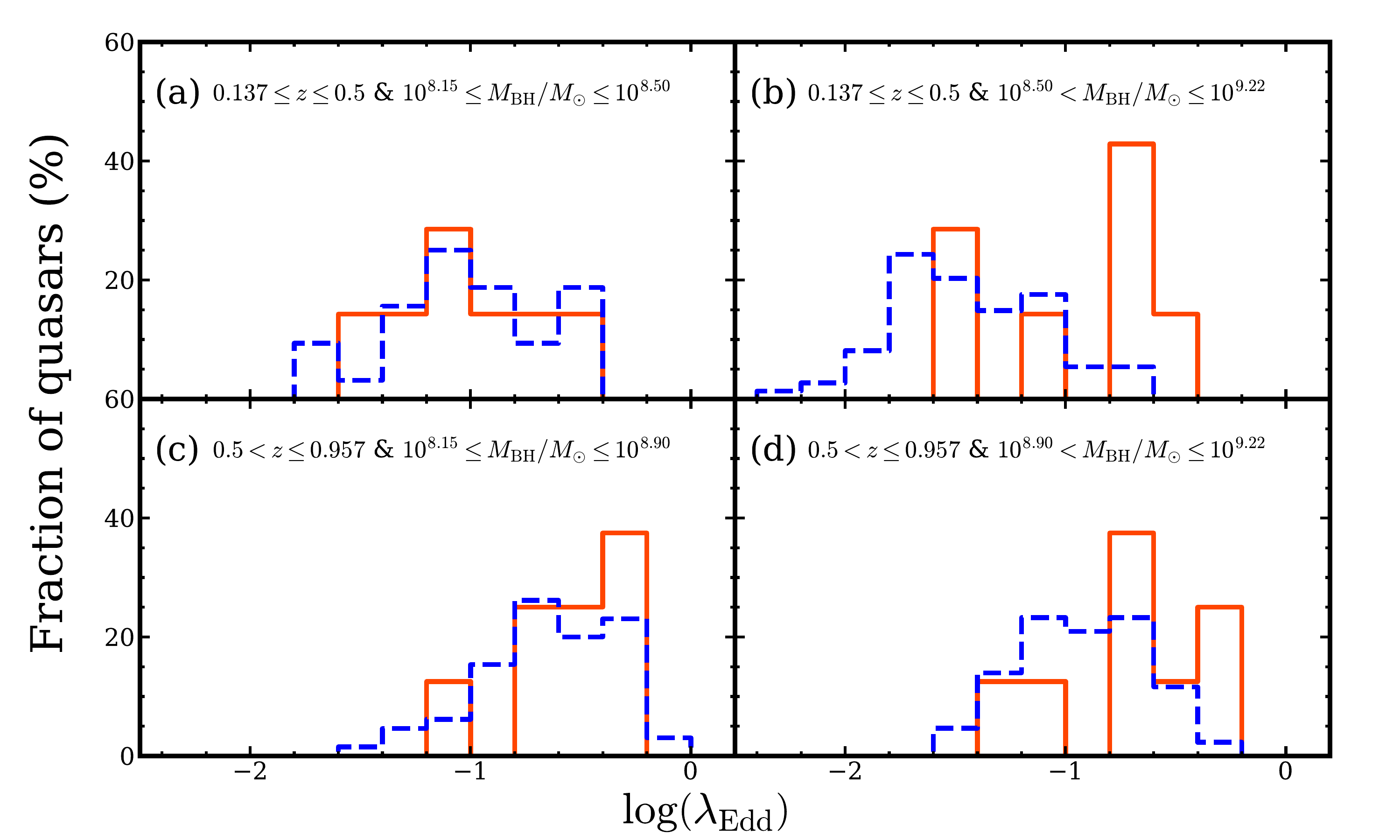}\\
		\caption{
			(a) $\lambda_{\rm Edd}$ distributions for low-$z$ and low-mass quasars.
			Red solid and blue dashed histograms denote the dust-obscured and the unobscured quasars, respectively.
			(b) $\lambda_{\rm Edd}$ distributions for low-$z$ and high-mass quasars.
			(c) $\lambda_{\rm Edd}$ distributions for high-$z$ and low-mass quasars.
			(d) $\lambda_{\rm Edd}$ distributions for high-$z$ and high-mass quasars.
			\label{fig:Edd-bins}}
	\end{figure*}

	Furthermore, given that $\lambda_{\rm Edd}$ can be affected by BH mass (e.g., \citealt{shen11,kim15a}),
	we also compare the $\lambda_{\rm Edd}$ distributions of four sub-samples divided
	by BH mass and redshift to mitigate the effects of BH mass and redshift on the $\lambda_{\rm Edd}$ distribution.
	First, we divide the dust-obscured quasars and the unobscured quasars into
	low-$z$ ($0.137 \leq z \leq 0.5$) and high-$z$ ($0.5 < z \leq 0.957$) samples.
	For the low-$z$ sample, we divide them into low-$z$ and low-mass ($10^{8.15}\,M_{\odot} \leq M_{\rm BH} \leq 10^{8.50}\,M_{\odot}$) and 
	low-$z$ and high-mass ($10^{8.50}\,M_{\odot} < M_{\rm BH} \leq 10^{9.22}\,M_{\odot}$) sub-samples.
	For the low-$z$ and low-mass sub-sample, 7 dust-obscured quasars and 32 unobscured quasars are selected,
	and their $\log \left( \lambda_{\rm Edd} \right)$ values are $-$0.99$\pm$0.31 and $-$0.94$\pm$0.36, respectively.
	The $\lambda_{\rm Edd}$ values of the dust-obscured and unobscured quasars are similar,
	with a considerable overlap due to large uncertainties.
	However, in other sub-samples, more significant differences are found.
	For the low-$z$ and high-mass sub-sample, 7 dust-obscured quasars and 74 unobscured quasars are selected,
	and their $\log \left( \lambda_{\rm Edd} \right)$ values are $-$0.61$\pm$0.38 and $-$1.33$\pm$0.35, respectively.
	After that, for the high-$z$ sample, we also divide them into high-$z$ and low-mass ($10^{8.15}\,M_{\odot} \leq M_{\rm BH} \leq 10^{8.90}\,M_{\odot}$) and
	high-$z$ and high-mass ($10^{8.90}\,M_{\odot} < M_{\rm BH} \leq 10^{9.22}\,M_{\odot}$) sub-samples.
	For the high-$z$ and low-mass sub-sample, 8 dust-obscured quasars and 65 unobscured quasars are chosen,
	and their $\log \left( \lambda_{\rm Edd} \right)$ values are $-$0.38$\pm$0.24 and $-$0.52$\pm$0.31, respectively.
	Moreover, for the high-$z$ and high-mass sub-sample, we obtain 8 dust-obscured quasars and 43 unobscured quasars,
	and their $\log \left( \lambda_{\rm Edd} \right)$ values are $-$0.56$\pm$0.36 and $-$0.82$\pm$0.28, respectively.
	Overall, the $\lambda_{\rm Edd}$ of the dust-obscured quasars are significantly higher than 
	those of the unobscured quasars in the limited redshift and $M_{\rm BH}$ range, 
	and these comparisons are shown in Figure \ref{fig:Edd-bins}.
	
	\section{Discussion}\label{sec:dis}
	\subsection{Comparisons of $L_{\rm bol}$ and $M_{\rm BH}$ Values of Dust-obscured Quasars from Different Estimators}\label{sec:Pb}
	The $L_{\rm bol}$ and $M_{\rm BH}$ values of dust-obscured quasars have been measured using various methods \citep{kim18a,urrutia12}.
	In this subsection, we examine the consistency between the $L_{\rm bol}$ and $M_{\rm BH}$ values measured with the $L_{\rm MIR}$-based estimators
	and those obtained from the different methods used in \cite{kim18a} and \cite{urrutia12}.
	
	Nine and four of our dust-obscured quasars are overlapped with 
	the dust-obscured quasars studied in \cite{kim18a} and \cite{urrutia12}, respectively.
	Using these objects, we compare the $L_{\rm bol}$ and $M_{\rm BH}$ values from our method versus those from other methods.
	First, for the nine dust-obscured quasars, 
	we compare their $L_{\rm bol}$ and $M_{\rm BH}$ values from the $L_{\rm MIR}$-based estimators
	to those from Paschen line-based estimators \citep{kim10,kim22}.
	\cite{kim18a} measured their P$\beta$ or P$\alpha$ line properties, luminosity and FWHM,
	and their $L_{\rm bol}$ and $M_{\rm BH}$ values can be measured
	with the Paschen line-based estimators established in \cite{kim22} and \cite{kim10}, respectively.
	The Paschen line-based $L_{\rm bol}$ estimators \citep{kim22} are
	\begin{equation}
		\begin{aligned}
			\log \left( \frac{L_{\rm bol}}{\rm 10^{42}\,erg~s^{-1}} \right) =& \left( 1.31 \pm 0.09 \right) \\
			& + \left( 0.97 \pm 0.05 \right) \log \left( \frac{L_{\rm P\beta}}{\rm 10^{40}\,erg~s^{-1}} \right)
		\end{aligned}
	\end{equation}
	and
	\begin{equation}
		\begin{aligned}
			\log \left( \frac{L_{\rm bol}}{\rm 10^{42}\,erg~s^{-1}} \right) =& \left( 1.43 \pm 0.11 \right) \\
			& + \left( 0.87 \pm 0.06 \right) \log \left( \frac{L_{\rm P\alpha}}{\rm 10^{40}\,erg~s^{-1}} \right).
		\end{aligned}
	\end{equation}
	Moreover, the Paschen line-based $M_{\rm BH}$ estimators \citep{kim10} are
	\begin{equation}
		\begin{aligned}
			\log \left( \frac{M_{\rm BH}}{M_{\odot}} \right) = & \left( 7.33 \pm 0.10 \right) + \left( 0.45 \pm 0.03 \right) \log \left( \frac{L_{\rm P\beta}}{\rm 10^{42}\,erg~s^{-1}} \right) \\
			& + \left( 1.69 \pm 0.16 \right) \log \left( \frac{\rm FWHM_{P\beta}}{\rm 10^{3}\,km~s^{-1}} \right)
		\end{aligned}
	\end{equation}
	and
	\begin{equation}
		\begin{aligned}
			\log \left( \frac{M_{\rm BH}}{M_{\odot}} \right) = & \left( 7.29 \pm 0.10 \right) + \left( 0.43 \pm 0.03 \right) \log \left( \frac{L_{\rm P\alpha}}{\rm 10^{42}\,erg~s^{-1}} \right) \\
			& + \left( 1.92 \pm 0.18 \right) \log \left( \frac{\rm FWHM_{P\alpha}}{\rm 10^{3}\,km~s^{-1}} \right).
		\end{aligned}
	\end{equation}
	Note that the FWHM and luminosity of both the P$\beta$ and P$\alpha$ lines for 1307$+$2338 were measured in \cite{kim18a},
	and its $L_{\rm bol}$ and $M_{\rm BH}$ values from both lines are used in this comparison.
	The measured $L_{\rm bol}$ and $M_{\rm BH}$ values based on the Paschen lines are summarized in Table \ref{tbl:prop_diffest}.
	
	Second, \cite{urrutia12} measured the monochromatic luminosities, $\lambda L_{\lambda}$, at 15\,$\mu$m (hereafter, $L_{\rm 15}$) 
	and FWHM values of Balmer lines for the four dust-obscured quasars.
	Their $L_{\rm 15}$ values can be converted to the $L_{\rm 4.6}$ values
	by applying the ratio of $L_{\rm 15}$ to $L_{\rm 4.6}$ in the quasar template of \cite{krawczyk13},
	and the $L_{\rm bol}$ values can be measured with Equation \ref{eqn:Lbol_46} using the converted $L_{\rm 4.6}$ values.
	Additionally, the $M_{\rm BH}$ values can be measured with Equations \ref{eqn:MBH_46_Hb} and \ref{eqn:MBH_46_Ha}
	after converting the $L_{\rm 15}$ values to $L_{\rm 4.6}$ values using the ratio.
	The measured $L_{\rm bol}$ and $M_{\rm BH}$ values from the $L_{\rm 15}$ are listed in Table \ref{tbl:prop_diffest}.
	
	\begin{table*}
		\centering
		\caption{BH properties of dust-obscured quasars from different estimators \label{tbl:prop_diffest}}
		\begin{tabular}{ccccccccc}
			\hline\hline
			\noalign{\smallskip}
			object name&	&	\multicolumn{3}{c}{$\log \left( L_{\rm bol} / {\rm erg~s^{-1}} \right)$}&	&
			\multicolumn{3}{c}{$\log \left( M_{\rm BH} / M_{\odot}\right)$}\\
			\cline{3-5} \cline{7-9}\\
			&	& P$\beta$&	P$\alpha$&	$L_{\rm 15}$\tablefootmark{a}&	& P$\beta$& P$\alpha$& $L_{\rm 15}$\\
			\noalign{\smallskip}
			\hline
			\noalign{\smallskip}
			
			F2M0036$-$0113&		&	45.46$\pm$0.22&		--&							--&			&	7.85$\pm$0.19&	--&							--\\
			F2M0729$+$3336&		&	--&								--&							46.55&	&	--&							--&							8.70$\pm$0.13\tablefootmark{b}\\
			F2M0817$+$4354&		&	--&								45.14$\pm$0.23&	--&			&	--&							8.24$\pm$0.28&	--\\
			F2M0830$+$3759&		&	--&								--&							46.05&	&	--&							--&							8.69$\pm$0.11\tablefootmark{b}\\
			F2M0841$+$3604&		&	--&								--&							45.95&	&	--&							--&							8.65$\pm$0.26\tablefootmark{c}\\
			F2M1113$+$1244&		&	46.49$\pm$0.28&		--&							46.84&	&	8.43$\pm$0.24&	--&							8.64$\pm$0.18\tablefootmark{b}\\
			F2M1209$-$0107&		&	46.21$\pm$0.26&		--&							--&			&	8.87$\pm$0.29&	--&							--\\
			F2M1227$+$5053&		&	46.20$\pm$0.32&		--&							--&			&	8.52$\pm$0.36&	--&							--\\
			F2M1307$+$2338&		&	45.24$\pm$0.21&		45.53$\pm$0.25&	--&			&	7.79$\pm$0.20&	8.06$\pm$0.24&	--\\
			F2M1313$+$1453&		&	46.30$\pm$0.25&		--&							--&			&	8.79$\pm$0.25&	--&							--\\
			F2M1434$+$0935&		&	46.21$\pm$0.25&		--&							--&			&	8.02$\pm$0.18&	--&							--\\
			F2M2325$-$1052&		&	46.12$\pm$0.37&		--&							--&			&	8.48$\pm$0.46&	--&							--\\
			\noalign{\smallskip}
			\hline
		\end{tabular}
	\tablefoot{\\
		\tablefoottext{a}{The uncertainty for $L_{\rm 15}$ is not provided by \cite{urrutia12}.
		The uncertainty for $L_{\rm bol}$ presented in Figure \ref{fig:Comp_L_M} is estimated on an assumption that
		$L_{\rm 15}$'s uncertainty amounts to 10\,\% of its value.}\\
		\tablefoottext{b}{The $M_{\rm BH}$ values are measured based on $L_{\rm 15}$ values and H$\beta$ widths.}\\
		\tablefoottext{c}{The $M_{\rm BH}$ values are measured based on $L_{\rm 15}$ values and H$\alpha$ widths.}
	}
	\end{table*}
	
\begin{figure*}
		\centering
		\includegraphics[width=\textwidth]{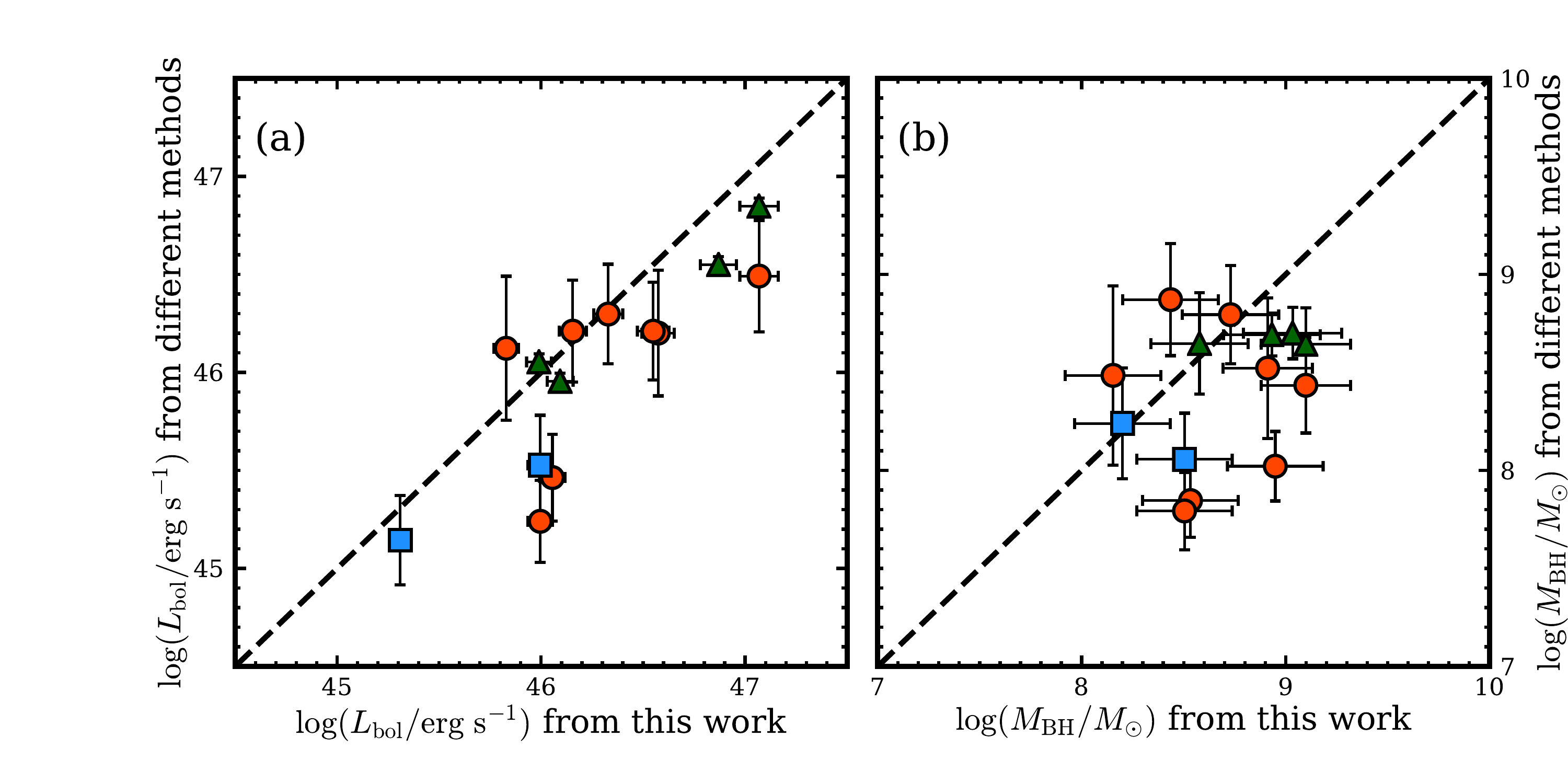}\\
		\caption{(a) Comparison of $L_{\rm bol}$ values from this work and different estimators.
			Red circles, blue squares, and green triangles denote
			the $L_{\rm bol}$ values are derived by using P$\beta$, P$\alpha$, and $L_{\rm 15}$, respectively.
			The dashed line denotes a line where the $L_{\rm bol}$ values are identical.
			(b) $M_{\rm BH}$ values from this work versus those from different methods. 
			Red circles and blue squares represent that the $M_{\rm BH}$ values are measured with P$\beta$ and P$\alpha$ lines, respectively.
			Green triangles denote the $M_{\rm BH}$ values are estimated from $L_{\rm 15}$ and Balmer line width.
			The dashed line represents a line where the $M_{\rm BH}$ values are identical.
			\label{fig:Comp_L_M}}
	\end{figure*}

	Figure \ref{fig:Comp_L_M} (a) presents a comparison of the $L_{\rm bol}$ values,
	showing a reasonable agreement between the two independent estimations,
	and their Pearson correlation coefficient is 0.831.
	Overall, the $L_{\rm bol}$ values derived in this work are slightly overestimated compared to
	the $L_{\rm bol}$ values obtained through the different methods, yet there is no clear tendency.
	
	Furthermore, a comparison of the $M_{\rm BH}$ values is also presented in Figure \ref{fig:Comp_L_M} (b).
	Compared to the $L_{\rm bol}$ comparison, there is a far less substantial agreement between the $M_{\rm BH}$ values,
	and their Pearson correlation coefficient is only 0.257. 
	This weak agreement might be due to their limited $M_{\rm BH}$ range,
	but the impact of the less robust line widths between Paschen and Balmer lines (e.g., \citealt{landt13})
	as a contributing factor cannot be ruled out.
	
	\subsection{Validity of $L_{\rm MIR}$-based $L_{\rm bol}$ estimators up to $z = 2.5$}\label{sec:est-higz}
	In this subsection, we explore the applicability of the $L_{\rm MIR}$ estimators up to $z = 2.5$,
	which were established using only SDSS quasars at $z \lesssim 0.5$ \citep{kim23}.
	To determine whether there is any significant deviation between the $L_{\rm MIR}$-based $L_{\rm bol}$ values
	and those derived from other estimators up to $z = 2.5$,
	we compare the $L_{\rm bol}$ values from the $L_{\rm 4.6}$-based estimator to 
	those measured from UV- and optical-based estimators \citep{kim23,runnoe12} 
	that were established with continuum luminosities, $\lambda L_{\lambda}$, 
	at 3000\,$\rm \AA{}$ and 5100\,$\rm \AA{}$ (hereafter, L3000 and L5100, respectively).
	
\begin{figure*}
		\centering
		\includegraphics[width=\textwidth]{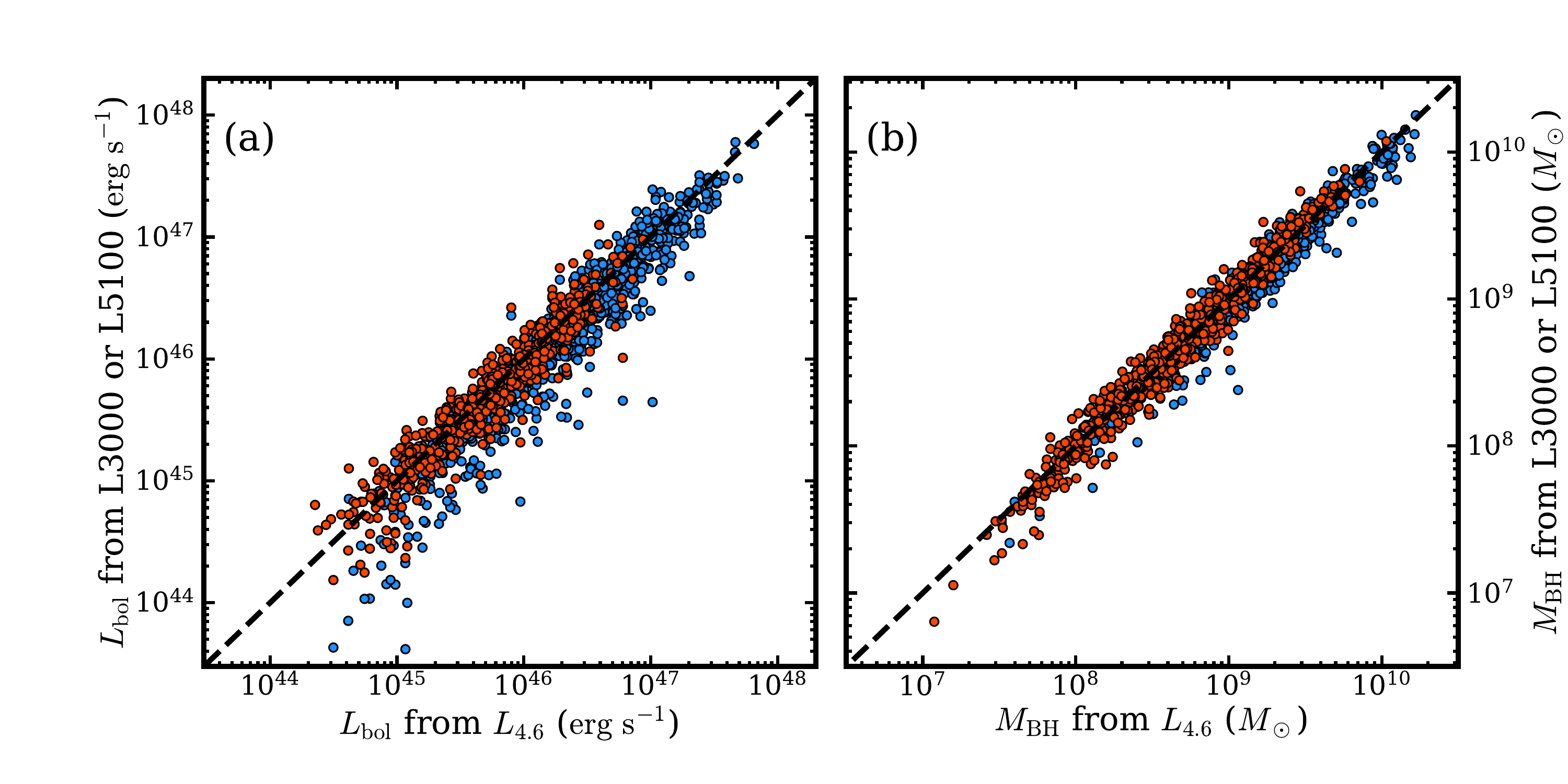}\\
		\caption{(a) Comparison of $L_{\rm bol}$ values from $L_{\rm 4.6}$ and those obtained using different estimators.
			Red circles mean the $L_{\rm bol}$ values derived from L5100, while blue circles denote those derived from L3000.
			The dashed line is a line where the $L_{\rm bol}$ values are identical.
			(b) $M_{\rm BH}$ values measured based on $L_{\rm 4.6}$ versus those from different methods. 
			Red circles represent the comparison of $M_{\rm BH}$ values measured using $L_{\rm 4.6}$ and $\rm FWHM_{H}$,
			compared to those derived from L5100 and $\rm FWHM_{H}$.
			Meanwhile, blue circles denote the $M_{\rm BH}$ values derived using $L_{\rm 4.6}$ and $\rm FWHM_{\ion{Mg}{II}}$,
			versus those from L3000 and $\rm FWHM_{\ion{Mg}{II}}$.
			The dashed line denotes a line where the $M_{\rm BH}$ values are identical.
			\label{fig:est-highz}}
	\end{figure*}

\begin{figure}
		\centering
		\includegraphics[width=\columnwidth]{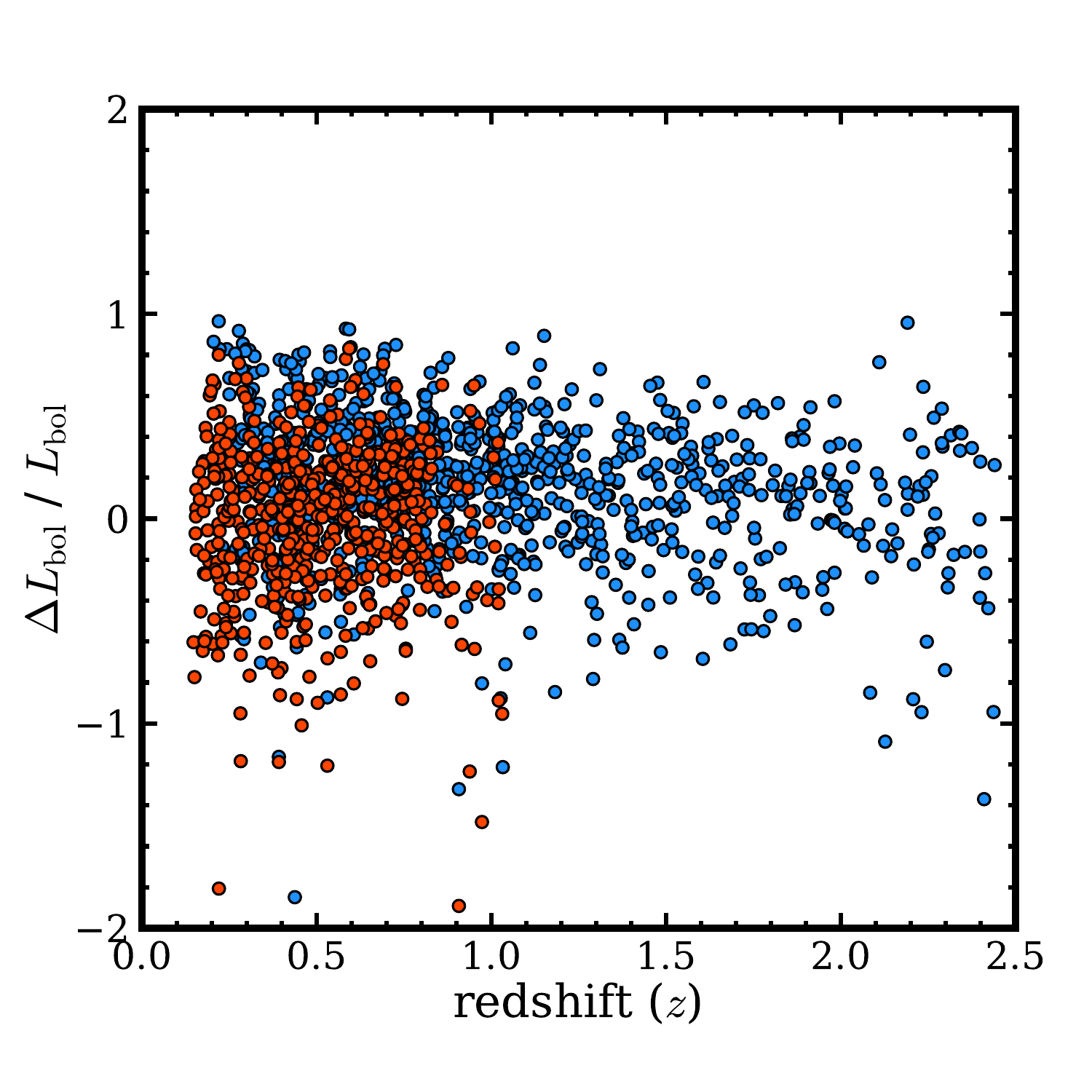}\\
		\caption{Redshift versus $\Delta L_{\rm bol} / L_{\rm bol}$,
			where $\Delta L_{\rm bol}$ means the discrepancy between $L_{\rm bol}$ measured from $L_{\rm 4.6}$
			and that from either L3000 or L5100.
			Red and blue circles denote that the $L_{\rm bol}$ values are derived from L5100 and L3000, respectively.
			\label{fig:est-highz-zdepen}}
	\end{figure}

	For the comparison, we select 1046 SDSS quasars at $z < 2.5$ among the SDSS DR14 quasars \citep{paris18},
	following the selection method for unobscured quasars described in Section \ref{sec:sample}.
	\cite{rakshit20} measured the L3000 and L5100 values for 959 and 563 SDSS quasars, respectively.
	Their $L_{\rm bol}$ values can be estimated with the L3000- and L5100-based $L_{\rm bol}$ estimators.
	The L3000-based $L_{\rm bol}$ estimator \citep{runnoe12} is
	\begin{equation}
		\begin{aligned}
			\log \left( \frac{L_{\rm bol}}{\rm 10^{44}\,erg~s^{-1}} \right) = & \left(0.53 \pm 1.27\right) \\
			& +\left( 0.97 \pm 0.03 \right) \log \left( \frac{\rm L3000}{\rm 10^{44}\,erg~s^{-1}} \right),
		\end{aligned}
	\end{equation}
	and the L5100-based $L_{\rm bol}$ estimator \citep{kim23} is
	\begin{equation}
		\begin{aligned}
			\log \left( \frac{L_{\rm bol}}{\rm 10^{44}\,erg~s^{-1}} \right) = & \left(0.764 \pm 0.019\right)\\
										& +\left( 0.978 \pm 0.029 \right) \log \left( \frac{\rm L5100}{\rm 10^{44}\,erg~s^{-1}} \right).
		\end{aligned}
	\end{equation}
	Figure \ref{fig:est-highz} (a) shows the comparison of $L_{\rm bol}$ derived from $L_{\rm 4.6}$ with those from L3000 and L5100,
	exhibiting a Pearson correlation coefficient of 0.952.
	Moreover, we investigate whether there is a redshift dependence for 
	the $L_{\rm bol}$ values measured from $L_{\rm 4.6}$ compared to those from other methods,
	as shown in Figure \ref{fig:est-highz-zdepen}. However, we found no evidence of redshift dependence.	
	Thus, although the $L_{\rm 4.6}$-based $L_{\rm bol}$ estimator was established using the SDSS quasars at $z \lesssim 0.5$,
	our results demonstrate that the $L_{\rm 4.6}$-based estimator accurately measures $L_{\rm bol}$ for quasars up to $z = 2.5$.
	
	Additionally, we compare $M_{\rm BH}$ derived from $L_{\rm 4.6}$ to those measured from other estimators.
	Among the 1046 SDSS quasars at $z < 2.5$, 
	\cite{rakshit20} measured $M_{\rm BH}$ values for 190 and 557 SDSS quasars
	using the L5100 combined with $\rm FWHM_{H\alpha}$ and $\rm FWHM_{H\beta}$, respectively.
	Furthermore, \cite{rakshit20} also estimated $M_{\rm BH}$ values for 804 SDSS quasars using L3000 and $\rm FWHM_{\ion{Mg}{II}}$.
	For the comparison, we determine $M_{\rm BH}$ values based on $L_{\rm 4.6}$ using Equation \ref{eqn:MBH_46_Hb} or \ref{eqn:MBH_46_Ha}
	for objects with measurements of $\rm FWHM_{H\beta}$ or $\rm FWHM_{H\alpha}$, respectively.
	For objects with $\rm FWHM_{\ion{Mg}{II}}$ measurements,
	we estimate their $M_{\rm BH}$ values using Equation \ref{eqn:MBH_46_Hb} 
	after converting $\rm FWHM_{\ion{Mg}{II}}$ to $\rm FWHM_{H\beta}$,
	based on the relationship described by \cite{bisogni17}.
	Figure \ref{fig:est-highz} (b) presents the comparison of $M_{\rm BH}$ derived from $L_{\rm 4.6}$ against those from L3000 and L5100,
	demonstrating the reliability of the $L_{\rm 4.6}$-based $M_{\rm BH}$ estimators up to $z = 2.5$.
	The Pearson correlation coefficient between the two quantities is 0.984.
	Note that in this comparison, we use the FWHM values measured from the same line,
	thus not accounting for the relationships and distributions between FWHM values of different lines (e.g., \citealt{kim10,landt13,bisogni17}).
	
	Although only some dust-obscured quasars at high-$z$ have been found to date (e.g., \citealt{banerji15}),
	it is expected that a significant number of high-$z$ dust-obscured quasars will be discovered
	through future IR surveys, such as Spectro-Photometer for the History of the Universe, Epoch of Reionization, and Ices Explorer (SPHEREx; \citealt{dore14}).
	As a result, the $L_{\rm MIR}$-based estimators are expected to play an important role for 
	unveiling the intrinsic properties of these high-$z$ dust-obscured quasars.

	\section{Conclusions}
	In the merger-driven galaxy evolution scenario as presented in Figure \ref{fig:Gal_evol}, 
	dust-obscured quasars are expected to occur as the intermediate stage galaxies between ULIRGs and unobscured quasars.
	This scenario is supported by the higher $\lambda_{\rm Edd}$ values of dust-obscured quasars
	compared to those of unobscured quasars \citep{kim15a,kim18a}.
	Although the BH activities of dust-obscured quasars are obscured,
	the previous studies used IR spectroscopic data to minimize the effects of dust extinction.
	However, the previous studies used the limited number of samples owing to
	(i) considerable expense associated with obtaining the IR spectroscopic data;
	(ii) the limited IR wavelength window in the atmosphere. 
	As a result, only a limited number of dust-obscured quasars have been studied,
	yielding statistics that are not robust and confined to narrow redshift ranges.
	
	In order to overcome these limitations of the previous studies,
	we estimated the $\lambda_{\rm Edd}$ values of 30 dust-obscured quasars at $z \lesssim 1$
	using the $L_{\rm 4.6}$-based $L_{\rm bol}$ and $M_{\rm BH}$ estimators \citep{kim23}
	that have the advantages of being widely applicable.
	Compared to the control sample of unobscured quasars,
	we found that the $\lambda_{\rm Edd}$ values of the dust-obscured quasars
	are significantly higher than those of the unobscured ones.
	This result was consistent across the wide redshift range ($z \lesssim 1$),
	and identical outcomes were yielded even when the analysis was constrained by $M_{\rm BH}$ or redshift. 
	
	Including our results, several observational studies have found that
	dust-obscured quasars have (i) high $\lambda_{\rm Edd}$ values \citep{urrutia12,kim15a,kim18b},
	(ii) high fractions of merging features in their host galaxies \citep{urrutia08,glikman15},
	(iii) dusty red colors \citep{kim18a}, (iv) merging SMBH systems \citep{kim20},
	and (v) enhanced star-formation activities \citep{georgakakis09}.
	These findings strongly support the picture that 
	dust-obscured quasars are the intermediate stage galaxies between ULIRGs and unobscured quasars,
	as outlined in Figure \ref{fig:Gal_evol}.
	
	
	\begin{acknowledgements}
		D.K. acknowledges the support of the National Research Foundation of Korea (NRF) 
		grant (Nos. 2021R1C1C1013580 and 2022R1A4A3031306) funded by the Korean Government (MSIT).
		Y. K. was supported by the National Research Foundation of Korea (NRF) grant funded by the Korean government (MSIT) (No. 2021R1C1C2091550).
		M.I. acknowledge the support from the National Research Foundation of Korea (NRF) grants, 
		No. 2020R1A2C3011091, and No. 2021M3F7A1084525, funded by the Korea government (MSIT).
		M.K. acknowledges the support by the National Research Foundation of Korea (NRF) grant (No. 2022R1A4A3031306).
		G.L. acknowledges support from the Basic Science Research Program through NRF funded by MSIT (No.2022R1A6A3A01085930).
	\end{acknowledgements}
	
	%
	%

\end{document}